\documentclass[a4paper,fleqn]{cas-sc}
\usepackage[numbers]{natbib}
\usepackage{booktabs} 
\usepackage{amsmath}  
\usepackage{hyperref}
\usepackage{doi}
\usepackage{tabularx}
\usepackage{makecell}
\usepackage{adjustbox}
\usepackage{placeins}
\usepackage{array}
\newcolumntype{Y}{>{\raggedright\arraybackslash}X}
\begin{document}
\let\WriteBookmarks\relax
\def\floatpagepagefraction{1}
\def\textpagefraction{.001}
\title
[mode=title]
{Full-gap kinetic limitation of thermionic-electron transport for electron transpiration cooling}

\shorttitle{Full-gap kinetic limitation of emitted-electron transport}
\shortauthors{Zhang et al.}

\author[1,2]{\textcolor{black}{Wushun Zhang}}
\author[1,2]{\textcolor{black}{Weixing Zhou}}
\author[1]{\textcolor{black}{Yinjian Zhao}}
\cormark[1]
\ead{zhaoyinjian@hit.edu.cn}
\affiliation[1]{organization={School of Energy 
Science and Engineering, 
Harbin Institute of Technology},
            city={Harbin},
            state={Heilongjiang 150001},
            country={PR China}}
\affiliation[2]{organization={Zhengzhou Research 
Institute of Harbin Institute of Technology},
                city={Zhengzhou},
                state={Henan 450000},
                country={PR China}}
\cortext[cor1]{Corresponding author}
\begin{abstract}
Electron transpiration cooling (ETC) can mitigate the aerothermal load on sharp hypersonic leading edges, 
but its effectiveness depends on whether thermionically emitted electrons can escape from the hot surface rather than return to it. 
In this work, a one-dimensional-in-space, three-dimensional-in-velocity electrostatic particle-in-cell/Monte Carlo collision model is developed to resolve ETC-relevant thermionic emission, collisional plasma transport, cathode-directed backflow, and downstream collection in a full cathode--anode plasma diode. 
A helium benchmark configuration is used as a controlled mechanism test for emitted-electron transport and backflow-limited current flow. 
As the imposed emission increases, the diode initially remains in a weak-backflow regime, in which the net emitted-electron transport and the downstream collected emitted-electron flux increase with emission. 
When the imposed emission is further increased, the cathode-directed backflow rises rapidly and a narrow transition to backflow-limited transport appears between emission fluxes of $7.0\times10^{19}$ and $7.5\times10^{19}\,\mathrm{m^{-2}\,s^{-1}}$. 
At $7.25\times10^{19}\,\mathrm{m^{-2}\,s^{-1}}$, the backflow ratio reaches $54.03\%$, and both the net transport efficiency and the downstream collection efficiency decrease to about $46\%$. 
Above this range, additional cathode-directed backflow overcompensates the imposed emission increase, so useful emitted-electron transport decreases rather than saturates. 
Boundary energy diagnostics show that stronger emission can continue to increase the nominal cathode-side boundary-cooling metric after the backflow transition, but this increase no longer corresponds to improved emitted-electron escape or full-gap transport. 
These results demonstrate that the present PIC-MCC framework can capture the key kinetic processes controlling ETC-relevant emitted-electron escape and backflow limitation.
\end{abstract}

\begin{keywords}
Electron transpiration cooling (ETC) \sep
Particle-in-cell (PIC) \sep
Monte Carlo collision (MCC) \sep
Thermionic emission \sep
Backflow-limited transport \sep
Full-gap plasma diode 
\end{keywords}

\maketitle

\section{Introduction}

Hypersonic vehicles are subjected to severe aerodynamic heating and wave drag during high-speed flight, making thermal protection and drag reduction central challenges in aerospace vehicle design \cite{zhangInvestigationHeatDrag2026}. 
This problem is especially critical near sharp leading edges, where geometric sharpening can reduce aerodynamic drag but also intensifies local aerothermal loading. 
Strong shock compression and viscous dissipation convert a large fraction of the freestream kinetic energy into internal energy in the shock layer, producing severe thermal loads on the vehicle surface. 
For air-breathing or long-duration hypersonic vehicles, aerodynamic efficiency usually requires sharp leading edges, whereas the stagnation-point heat flux increases rapidly as the leading-edge radius decreases. 
Classical analyses by Lees and Kubota and by Fay and Riddell established this fundamental conflict between aerodynamic performance and thermal protection \cite{leesInviscidHypersonicFlow1957,fay_theory_1958}. 
Therefore, thermal-protection strategies for hypersonic vehicles have been extensively investigated, including passive, semi-passive, and active approaches \cite{zhuReviewActiveThermal2018,zhang_research_2020}. 
Recent reviews and numerical studies further show that active flow-control and cooling concepts remain important for reusable sharp-edged vehicles operating under severe aerodynamic heating \cite{zhang_thermal_2025,luoDragReductionThermal2025}.

In addition to severe heating, atmospheric reentry and sustained hypersonic flight may generate an ionized plasma sheath around the vehicle surface. 
This near-wall plasma environment can modify charged-particle transport, surface charging, and plasma--surface energy exchange \cite{zhangAnalysisInfluence2022,guoRapidPrediction2025}. 
Plasma-based technologies have also been increasingly investigated for aerospace applications under wide-range operating environments, including plasma ignition, flow control, and other strongly coupled gas--plasma processes \cite{zhangStudyCharacterization2026}. 
For electron-emitting thermal-protection concepts, the plasma sheath is not only an electromagnetic environment but also a transport medium through which emitted electrons must escape from the hot wall. 
Therefore, the effectiveness of electron-based surface cooling depends on the coupled evolution of thermionic emission, plasma sheath structure, and charged-particle transport near the emitting surface.

Electron transpiration cooling (ETC) has emerged as a promising active thermal-protection concept for sharp hypersonic leading-edge structures. 
ETC is based on thermionic emission: when a high-temperature surface emits electrons, part of the surface thermal energy is converted into electron energy and transported away from the hot region. 
Early conceptual studies proposed ETC as a possible cooling mechanism for sharp hypersonic leading edges and hot aerospace surfaces \cite{alkandry_conceptual_2014,uribarri_electron_2015}. 
Subsequent modeling work incorporated thermionic emission into hypersonic flowfield simulations and showed that ETC can reduce leading-edge temperature under favorable material and flow conditions \cite{hanquist_modeling_2016,hanquist_detailed_2017}. 
Further studies examined the effectiveness of thermionic emission, plasma-assisted cooling, operation thresholds, and material or flowfield factors for ETC-based thermal protection systems \cite{hanquist_effectiveness_2018,liImpactingFactorsOperation2024}.

Recent ETC studies have moved beyond local emission models toward more complete interpretations of electron transport and energy redistribution. 
Flight-regime analyses have shown that the cooling performance of ETC depends on altitude, flight velocity, wall temperature, material work function, and the surrounding plasma state \cite{gibbons_2025,monroeInfluenceHypersonic2025}. 
Recent system-level and circuit-oriented studies further emphasize that ETC should not be treated as a local emission process alone, because the emitted electrons must be transported away from the hot surface and eventually returned through the vehicle structure \cite{monroeElectronTranspirationCircuits2025b,boyer_mechanisms_2025}. 
This system-level view is important because emitted electrons may either contribute to downstream energy redistribution or return to the cathode and reduce the useful cooling benefit.

A central difficulty in ETC is that thermionic emission does not directly translate into useful cross-gap current. 
The emitted current can be strongly modified by plasma sheaths, space-charge effects, and particle transport in the near-wall plasma. 
Classical emitting-sheath theories show that increasing emission weakens the sheath and may lead to space-charge-limited conditions or virtual-cathode formation \cite{hobbsHeatFlowLangmuir1967,takamuraSpaceChargeLimitedCurrent2004a}. 
Kinetic theory further demonstrates that the emitted-electron temperature and plasma-electron temperature can strongly affect the sheath potential near electron-emitting surfaces \cite{PhysRevLett.111.075002}. 
For ETC applications, sheath models have also been developed to account for thermionic emission, space-charge limitation, and the coupling between emitted current and plasma transport near hypersonic surfaces \cite{sahuPlasmaSheathModels2021,vatanseverParametricStudy2022}.

More recent studies have shown that strongly emitting plasma-facing surfaces may enter regimes that differ fundamentally from conventional space-charge-limited sheath pictures. 
Campanell and co-workers demonstrated that intense emission can produce inverse-sheath states in which the wall potential is above the nearby plasma potential and ion collection is strongly modified \cite{campanellNegativePlasmaPotential2013,PhysRevLett.116.085003}. 
Related studies further showed that hot-cathode plasma diodes can operate in inverse modes and that thermionic emission can alter plasma-surface energy exchange by cooling the target-side plasma electron population \cite{campanellAlternativeModelSpacechargelimited2018a,PhysRevLett.122.015003}. 
These results imply that the plasma-surface interaction in ETC cannot be fully understood from classical sheath theory alone, especially when emission is strong and the emitted electrons interact with the plasma over a finite transport distance.

Recent full-diode studies have further clarified that emitted-electron current limitation can be a global problem rather than a purely local cathode-sheath problem. 
Campanell et al. demonstrated that the emitted current from a cathode to an anode can be limited by two distinct mechanisms: space-charge limitation associated with virtual-cathode formation and backflow saturation associated with a weakened cathode sheath, plasma-interior transport, and anode-sheath limitation \cite{PhysRevLett.134.145301}. 
Depending on collisionality and emission level, these mechanisms may appear separately, cooperatively, or in a competing oscillatory form. 
This picture is highly relevant to ETC because an ETC system also requires thermionically emitted electrons to pass through a plasma-filled region and be collected downstream. 
Therefore, a full cathode--anode description is required to evaluate emitted-electron transmission, cathode-directed backflow loss, and the resulting limitation of useful ETC-relevant transport.

Although existing ETC studies have made substantial progress, many of them rely on continuum flowfield models, analytical sheath models, or prescribed emission boundary conditions. 
These approaches are valuable for system-level estimates, but they do not fully resolve the kinetic evolution of emitted electrons, plasma electrons, ions, self-consistent electric fields, and collisional scattering in the full transport gap. 
Particle-in-cell/Monte Carlo collision (PIC-MCC) methods provide a direct kinetic framework for this problem and have been widely used for plasma-discharge and plasma-surface-interaction simulations \cite{ck_birdsall_plasma_1991,hockneyComputerSimulationUsing}. 
In PIC-MCC simulations, charged-particle dynamics are solved self-consistently with the electrostatic field, while electron-neutral and ion-neutral collisions are treated probabilistically using cross-section data \cite{vahediMonteCarloCollision1995}. 
This makes PIC-MCC suitable for investigating how collisional isotropization, charge-exchange processes, excitation losses, sheath restructuring, and non-Maxwellian velocity distributions jointly affect thermionic-electron transport.

In this work, we develop a one-dimensional-in-space, three-dimensional-in-velocity (1D--3V) electrostatic PIC-MCC model of a full cathode--anode plasma diode for ETC using WarpX~\cite{warpx}. 
The helium collisional diode of Campanell et al.~\cite{PhysRevLett.134.145301} is adopted as a controlled mechanism benchmark rather than as a point-by-point curve-reproduction target. 
The aim is to test whether full-diode emitted-current limitation persists in an ETC-oriented stochastic kinetic description with Monte Carlo collisions and three-dimensional velocity-space redistribution. 
The emphasis is placed on the emitted-electron transport quantities that matter for ETC, including cathode escape, cathode-directed backflow, net emitted-electron transport, downstream collection, and cathode-side boundary energy diagnostics. 
The downstream electrode is treated as a one-dimensional collection boundary and transport diagnostic rather than as a finite-area collector in a vehicle configuration.

\section{Model}
\label{sec:model}

\subsection{Numerical setup}
\label{sec:numerical_setup}

\begin{figure}[htbp]
\centering
\includegraphics[width=0.78\textwidth]{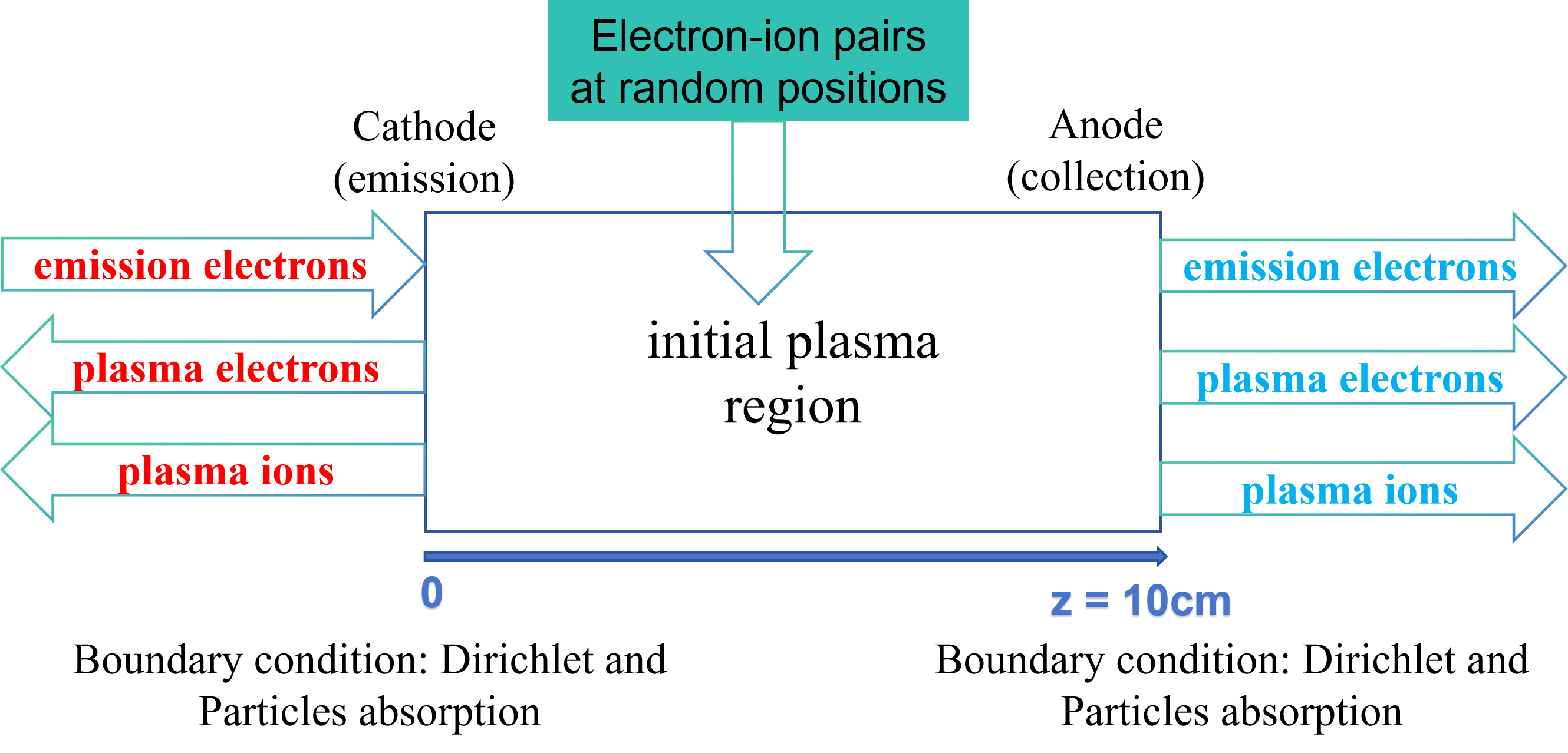}
\caption{
Schematic of the 1D--3V cathode--anode PIC-MCC model and computational domain.
The cathode at $z=0$ emits thermionic electrons, while particles reaching either electrode are absorbed.
}
\label{fig:model_schematic}
\end{figure}

The numerical setup follows 
the collisional benchmark configuration 
of Campanell \emph{et al.} \cite{PhysRevLett.134.145301}. 
The initial plasma number density 
is $n_{0}=5\times10^{14}\ \mathrm{m^{-3}}$. 
The initial plasma electron and ion temperatures 
are $T_{\mathrm{e},0}=3\ \mathrm{eV}$ and $T_{\mathrm{i},0}=0.15\ \mathrm{eV}$, 
respectively, 
and the emission electron temperature 
is $T_{\mathrm{c}}=0.2\ \mathrm{eV}$. 
The corresponding Debye length
of the background plasma 
is $\lambda_{\mathrm{D}}=(\varepsilon_{0}T_{\mathrm{e},0}/n_{0}e)^{1/2}\approx575.83\ \mu\mathrm{m}$, 
the electron thermal velocity 
is $v_{\mathrm{te}}=(eT_{\mathrm{e},0}/m_{\mathrm{e}})^{1/2}\approx7.26\times10^{5}\ \mathrm{m\,s^{-1}}$, 
the ion thermal velocity is $v_{\mathrm{ti}}=(eT_{\mathrm{i},0}/m_{\mathrm{i}})^{1/2}\approx1.90\times10^{3}\ \mathrm{m\,s^{-1}}$, 
and the emission electron thermal velocity 
is $(eT_{\mathrm{c}}/m_{\mathrm{e}})^{1/2}\approx1.88\times10^{5}\ \mathrm{m\,s^{-1}}$. 
For helium ions with mass $m_{\mathrm{i}}\approx6.64\times10^{-27}\ \mathrm{kg}$, 
the ion sound speed is $c_{\mathrm{s}}=(eT_{\mathrm{e},0}/m_{\mathrm{i}})^{1/2}\approx8.51\times10^{3}\ \mathrm{m\,s^{-1}}$.

As shown in Fig. \ref{fig:model_schematic},
the domain length is set to $L=10\ \mathrm{cm}$. 
The cathode is located at $z=0$ and the anode at $z=L$.
Dirichlet boundary conditions are imposed 
for the electrostatic potential, 
with $\phi(0)=0\ \mathrm{V}$ and $\phi(L)=40\ \mathrm{V}$. 
The electrostatic potential is solved self-consistently from the charge density at each time step, 
and the electric field is obtained from the spatial gradient of the potential. 
Thus, the sheath structure, 
virtual-cathode-like potential depression, 
and full-gap potential redistribution 
are not prescribed, 
but arise from the kinetic particle dynamics 
and boundary conditions.
The computational grid contains $N_{z}=4096$ cells, 
giving a cell size of 
$\Delta z=L/N_{z}\approx24.41\ \mu\mathrm{m} \approx0.04\lambda_{\mathrm{D}}$
to resolve the sheath structure accurately.
The time step is chosen as 
$\Delta t=0.2\Delta z/v_{\mathrm{te}}$, 
which gives $\omega_{\mathrm{pe}}\Delta t\approx0.008$, 
where $\omega_{\mathrm{pe}}=(n_{0}e^{2}/m_{\mathrm{e}}\varepsilon_{0})^{1/2}\approx1.26\times10^{9}\ \mathrm{rad\,s^{-1}}$. 
The number of macro-particles per cell is
$N_{\mathrm{ppc}}=1000$, 
corresponding to a one-dimensional planar macro-particle weight
$w_{0}=n_{0}\Delta z/N_{\mathrm{ppc}}\approx1.22\times10^{7}\ \mathrm{m^{-2}}$.
Here, $w_0$ represents the number of real particles per macro-particle per unit transverse area, because the simulation is one-dimensional in space but represents a planar diode.

Initially, 
electrons and ions are uniformly distributed 
over the whole domain 
with Maxwellian velocity distributions 
at $T_{\mathrm{e},0}$ and $T_{\mathrm{i},0}$, 
respectively, 
and zero drift velocity. 
Particles reaching either boundary 
are absorbed and removed from the simulation. 
To maintain the prescribed background plasma inventory during the long transient evolution, 
the particle-replenishment procedure follows Campanell et al.~\cite{PhysRevLett.134.145301}. 
At each time step, the total number of ions lost to the two electrodes is counted, 
and the same number of electron--ion pairs is re-injected uniformly in the volume 
with overlapping random positions and Maxwellian velocities.

\FloatBarrier

\subsection{Emission and collisions}
\label{sec:emission_collision}

\begin{table}[htbp]
\centering
\caption{
Compact mechanism-level correspondence between the short-mean-free-path diode benchmark of Campanell et al.~\cite{PhysRevLett.134.145301} and the present 1D--3V PIC-MCC model.
}
\label{tab:model_comparison_prl}
\footnotesize
\setlength{\tabcolsep}{3pt}
\renewcommand{\arraystretch}{1.05}
\begin{tabularx}{\textwidth}{>{\raggedright\arraybackslash}p{3.0cm}YY}
\toprule
Item & Campanell et al.~\cite{PhysRevLett.134.145301} & Present work \\
\midrule
Geometry and species 
& 1D planar He diode, $L_x=10~\mathrm{cm}$; electrons + He$^+$ 
& 1D planar He diode, $L=10~\mathrm{cm}$; electrons + He$^+$ \\

Initial state 
& $n_0=5\times10^{14}~\mathrm{m^{-3}}$, $T_{e,0}=3~\mathrm{eV}$, $T_{i,0}=0.15~\mathrm{eV}$, $T_{\mathrm{emit}}=0.2~\mathrm{eV}$ 
& same, with $T_{\mathrm c}=0.2~\mathrm{eV}$ \\

Boundary conditions 
& cathode $0~\mathrm{V}$, anode $40~\mathrm{V}$ 
& same \\

Emission treatment 
& imposed emitted-flux scan 
& imposed emitted-flux scan \\

Mean-free-path ratios 
& $e_{\mathrm{mfp}}/L_x=0.05$, $i_{\mathrm{mfp}}/L_x=0.2$ 
& $\lambda_e/L=0.05$, $\lambda_i/L=0.2$ \\

Collision implementation 
& effective continuum collisional relaxation 
& MCC with fixed effective He elastic and CX cross sections \\

Velocity space and solver 
& 1D--1V continuum kinetic diode model, $f(x,v_x)$ 
& 1D--3V electrostatic PIC-MCC in WarpX, $f(z,v_x,v_y,v_z)$ \\
\bottomrule
\end{tabularx}
\end{table}

A prescribed thermionic emission flux 
is imposed at the cathode boundary. 
This choice is made 
because the main benchmark quantity in Campanell et al.~\cite{PhysRevLett.134.145301} 
is the dependence of the transmitted current and backflow-limited transport state 
on the imposed emission level.
By directly controlling the injected flux, 
the transition from unsaturated transport 
to backflow-limited transport 
can be identified more clearly. 
The emitted macro-electron number per time step is determined from
$N_{\mathrm{emit}}=\Gamma_{\mathrm{emit}}\Delta t/w_0$ 
with stochastic treatment of the fractional part,
where $\Gamma_{\mathrm{emit}}$ is the emitted-electron flux.
Each emitted electron is sampled 
from a half-Maxwellian distribution 
corresponding to $T_{\mathrm{c}}$, 
and only particles moving 
from the cathode into the domain are retained. 
The initial position of each emitted electron 
is set to $z_{\mathrm{p}}=v_{z}\Delta t$, 
so that the particle has just entered 
the simulation domain after one time step.
For diagnostic purposes, 
emitted electrons and background plasma electrons 
are tracked as separate electron populations. 
This separation does not change the electrostatic field solve, 
because both populations carry the same electron charge 
and contribute to the total charge density, 
but it allows the emitted-electron backflow flux, 
net emitted-electron flux, 
and anode-collected emitted-electron flux 
to be evaluated separately.

Collisions with the helium background are modeled 
by the Monte Carlo collision (MCC) method 
\cite{vahediMonteCarloCollision1995}. 
In the benchmark configuration, 
only two channels are retained: 
electron--neutral elastic scattering 
and ion--neutral charge-exchange (CX) scattering. 
This choice follows the physical picture 
of Campanell \emph{et al.}~\cite{PhysRevLett.134.145301}, 
in which electron collisionality 
is the main control parameter for the saturation path, 
while the ion mean free path 
is kept fixed and ion collisions are represented by CX.

The collisional parameters are chosen to
reproduce the short-mean-free-path benchmark regime discussed 
by Campanell \emph{et al.}~\cite{PhysRevLett.134.145301}. 
In that study, 
the representative case uses 
$e_{\mathrm{mfp}}/L_x=0.05$ and $i_{\mathrm{mfp}}/L_x=0.2$. 
In the present MCC implementation, 
we follow this benchmark in 
an equivalent mean-free-path sense 
by using a uniform neutral background density
$N_{\mathrm{INERT}}=2.9\times10^{21}~\mathrm{m^{-3}}$, 
a fixed electron--neutral elastic cross section 
$\sigma_{e}=6.8877\times10^{-20}~\mathrm{m^2}$
for any electron energy, 
and a fixed ion--neutral charge-exchange cross section 
$\sigma_{i}=1.7241\times10^{-20}~\mathrm{m^2}$
for any ion energy. 
These values give 
$\lambda_{\mathrm e}=1/(N_{\mathrm{INERT}}\sigma_e)=0.50~\mathrm{cm}$ 
and $\lambda_{\mathrm i}=1/(N_{\mathrm{INERT}}\sigma_i)=2.0~\mathrm{cm}$,
corresponding to $\lambda_{\mathrm e}/L=0.05$ 
and $\lambda_{\mathrm i}/L=0.2$ for $L=10~\mathrm{cm}$.
These fixed cross sections are used as effective benchmark cross sections 
to reproduce the prescribed mean-free-path ratios, 
rather than as energy-dependent realistic helium data. 

Although the nominal electron and ion mean-free-path ratios 
are matched to the short-mean-free-path benchmark, 
the same values of $\lambda_e/L$ and $\lambda_i/L$ 
do not uniquely determine the collision operator. 
In the continuum diode model of Campanell et al.~\cite{PhysRevLett.134.145301}, 
electron collisionality is represented through an effective one-dimensional relaxation scale. 
In the present 1D--3V PIC-MCC model, 
by contrast, electron--neutral elastic collisions are treated 
as stochastic velocity-space scattering events. 
Therefore, an electron collision can redistribute axial momentum 
into transverse velocity components, 
changing the effective axial mobility, residence time, 
and probability of cathode-directed return of emitted electrons. 
This distinction should be kept in mind when comparing 
the present high-emission branch 
with the representative continuum plateau-like branch.

In the MCC operator, 
the prescribed neutral temperature defines 
the velocity distribution of the neutral collision partner 
and is not a post-collision electron temperature. 
For electron--neutral elastic scattering, 
the neutral target temperature is set to 
$T_{n,e}=300~\mathrm{K}$, 
representing a room-temperature helium background. 
Because the helium atom is much heavier than the electron, 
elastic electron--neutral collisions mainly randomize 
the electron direction and only weakly exchange kinetic energy 
with the neutral bath. 
Therefore, this setting should not be interpreted 
as forcing the electron population toward 
a 300 K Maxwellian distribution. 
For ion--neutral charge exchange, 
the neutral target temperature is set to 
$T_{n,i}=1740~\mathrm{K}$, 
corresponding approximately to the initial ion temperature 
$T_{i,0}=0.15~\mathrm{eV}$. 
A selected comparison case with $T_{n,e}=1740~\mathrm{K}$ 
is included in Sec.~3.4 to show that 
the electron-side neutral-temperature choice 
has only a minor quantitative effect on the emitted-electron transport behavior.

For clarity, 
Table~\ref{tab:model_comparison_prl} summarizes 
the correspondence between the short-mean-free-path benchmark 
of Campanell \emph{et al.}~\cite{PhysRevLett.134.145301} 
and the present 1D--3V PIC-MCC model.
To isolate the roles of different collisional mechanisms,
additional comparison cases 
are considered beyond the benchmark configuration. 
These include elastic-plus-excitation cases 
and a selected case with ion CX disabled. 
Since all other numerical parameters 
are kept unchanged, 
the separate influences of inelastic electron energy loss 
and ion collisional trapping on sheath formation 
and backflow-limited transport 
can be identified more clearly.
For the excitation-on comparison cases, 
electron-impact excitation is included 
as an additional inelastic electron--neutral collision channel 
using the same helium excitation cross-section dataset 
as in the implemented MCC database. 
The excitation cases are used only as selected sensitivity tests 
and are not included in the baseline benchmark scan.

\subsection{Benchmark cases}
\label{sec:benchmark_validation}

The present model is assessed 
using a benchmark suite rather than 
a single operating point. 
All benchmark cases are performed 
under the same plasma and numerical conditions, 
while only the prescribed emission flux and, 
for selected cases, 
the activated collision channels are varied. 
The benchmark suite includes
$\Gamma_{\mathrm{emit}}=1.5\times10^{19}$, 
$2.0\times10^{19}$, 
$4.5\times10^{19}$, 
$5.0\times10^{19}$, 
$6.0\times10^{19}$, 
$7.0\times10^{19}$, 
$7.25\times10^{19}$, 
$7.5\times10^{19}$, 
$8.0\times10^{19}$, 
$1.0\times10^{20}$, 
and $1.2\times10^{20}\,\mathrm{m^{-2}\,s^{-1}}$.

The net emitted electron flux is evaluated as
$\Gamma_{\mathrm{net}}=\Gamma_{\mathrm{emit}}-\Gamma_{\mathrm{ref}}$,
where $\Gamma_{\mathrm{ref}}$ denotes the backflow flux 
of emitted electrons returning to the cathode. 
The backflow ratio is defined as
$\Gamma_{\mathrm{ref}}/\Gamma_{\mathrm{emit}}\times100\%$.
The anode-collected emitted-electron flux, 
$\Gamma_{\mathrm{anode}}$, 
is monitored independently as a cross-gap transport diagnostic.
The reported fluxes are averaged over selected late-time intervals in which the boundary fluxes fluctuate around statistically stable mean values.

For selected emission levels, 
additional comparisons are made between cases 
with and without excitation, 
and between cases with and without 
ion charge-exchange collisions. 
In this way, 
the benchmark suite covers low-, intermediate-, and high-emission regimes,
from weak-backflow transport to strongly backflow-limited transport.
The model is assessed from three complementary aspects: 
(i) the dependence of the anode-collected flux on the imposed emission level, 
(ii) the evolution of the full-gap potential profile, 
and (iii) 
the formation of near-cathode virtual-cathode-like structures and
full-gap potential collapse in the high-flux regime.

\subsection{Boundary energy diagnostics}
\label{sec:boundary_energy}

To relate the emitted-electron transport state to ETC-relevant thermal performance, 
boundary energy fluxes are evaluated as post-processed diagnostics for the corresponding late-time transport states. 
The purpose of this analysis is not to construct a complete vehicle-level heat balance, 
but to quantify the nominal energy removed from the emitting cathode, 
the energy returned to the cathode by emitted-electron backflow and incident plasma particles, 
and the charged-particle energy reaching the downstream collection boundary in the present benchmark diode.

This boundary-energy bookkeeping follows the ETC-circuit perspective, 
in which a complete ETC system involves not only an emitting cathode, 
but also a downstream collection boundary and an electron return path through the vehicle structure 
\cite{monroeElectronTranspirationCircuits2025b}. 
In the present one-dimensional model, however, the downstream boundary is used only as a collection and transport diagnostic. 
It should not be interpreted as a finite-area downstream collector surface in an actual hypersonic vehicle configuration.

\begin{figure*}[htbp]
\centering
\begin{minipage}{0.48\textwidth}
    \centering
    \includegraphics[width=\textwidth]{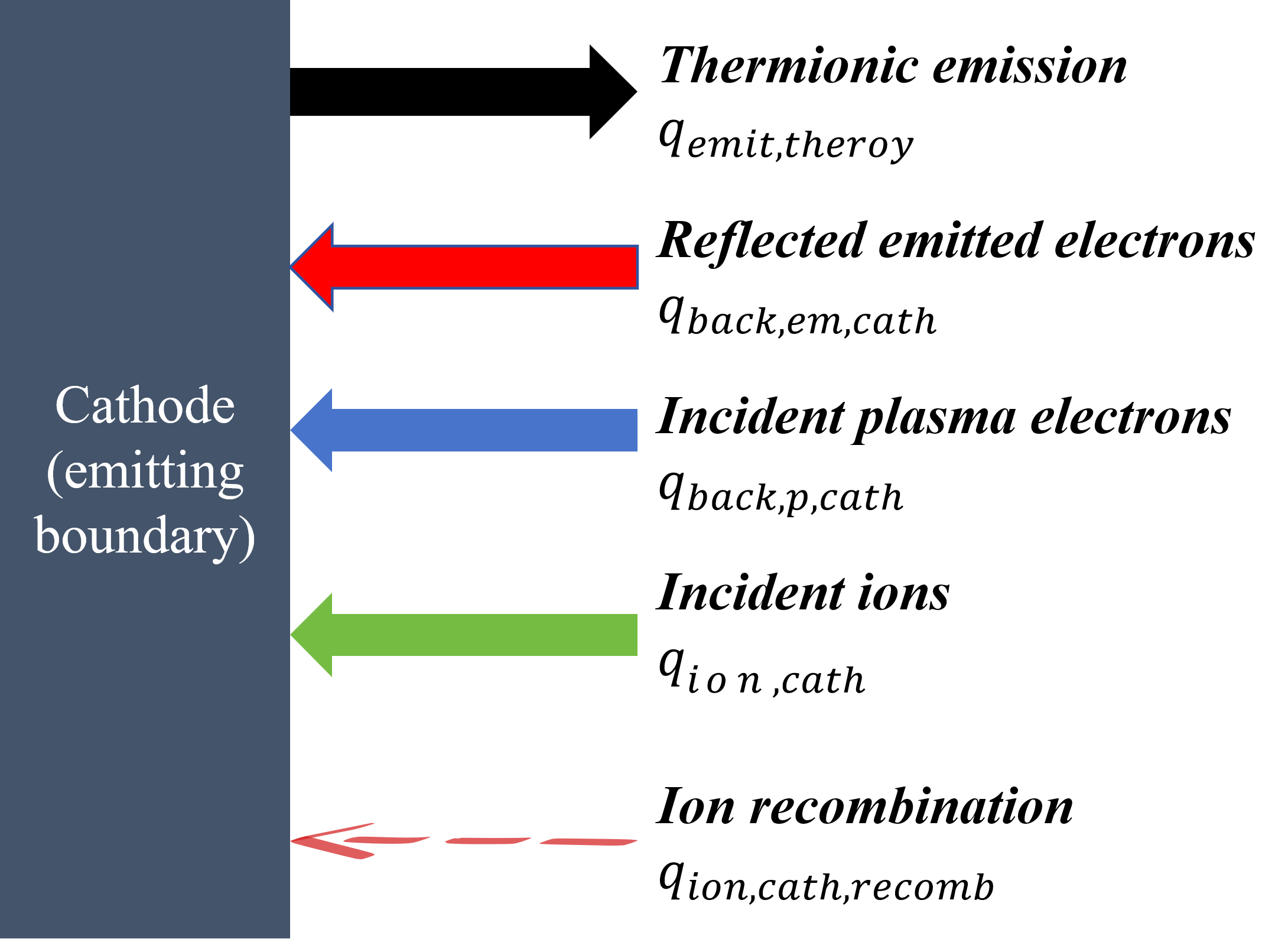}
    \vspace{1mm}
    (a)
\end{minipage}
\hfill
\begin{minipage}{0.48\textwidth}
    \centering
    \includegraphics[width=\textwidth]{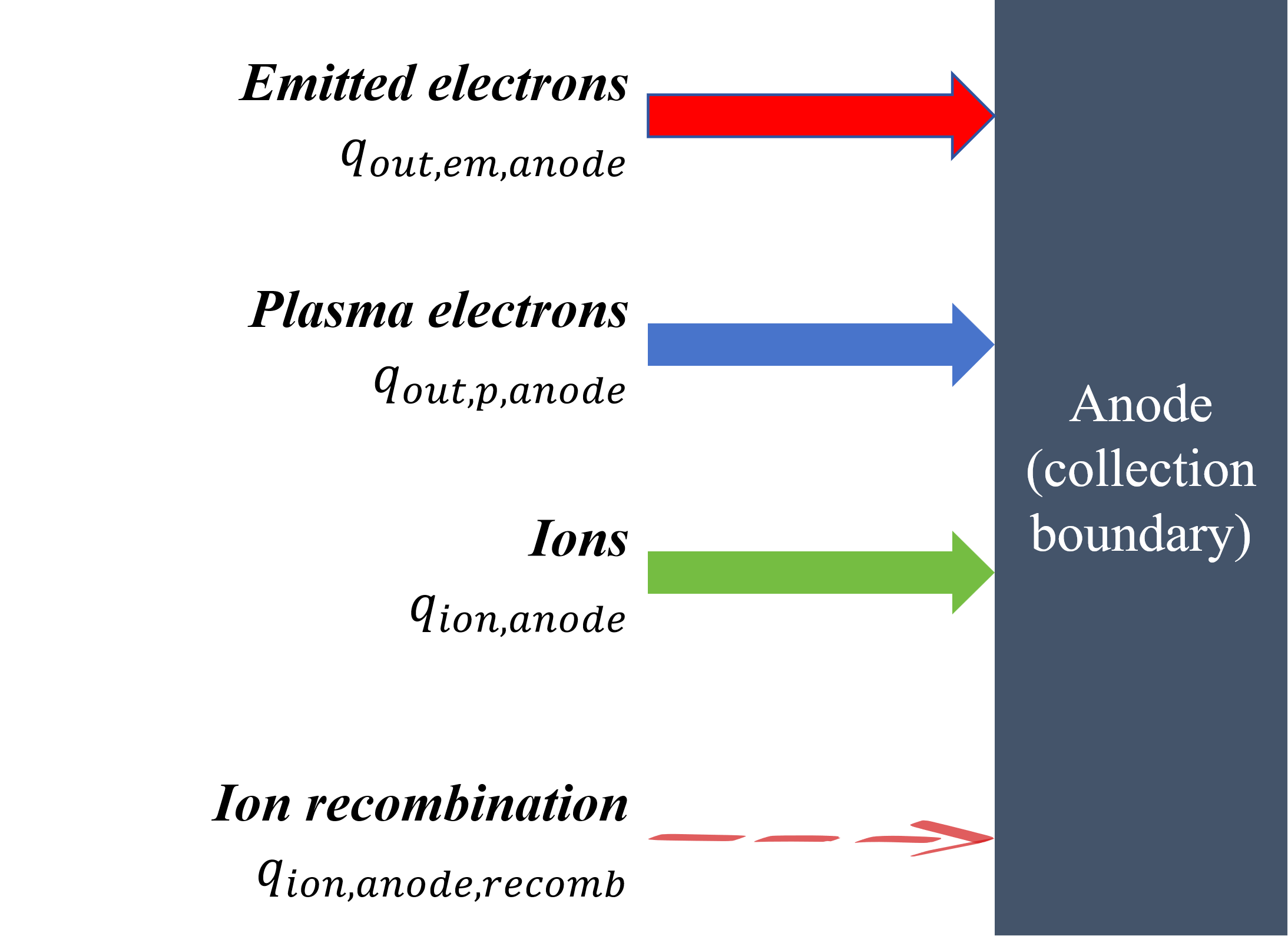}
    \vspace{1mm}
    (b)
\end{minipage}
\caption{
Schematic of boundary energy and flux accounting in the one-dimensional cathode--anode plasma diode:
(a) cathode-side emitted-electron cooling and returned particle-energy deposition;
(b) downstream transmitted-energy diagnostics at the collection boundary.
The cathode boundary emits thermionic electrons and receives cathode-directed backflow emitted electrons, incident plasma electrons, and incident ions.
The downstream boundary collects emitted electrons, plasma electrons, and ions transmitted across the plasma gap.
The downstream-side terms are used only as one-dimensional transmitted-energy diagnostics.
}
\label{fig:boundary_energy_schematic}
\end{figure*}

A schematic illustration of the boundary energy and flux accounting is shown in Fig.~\ref{fig:boundary_energy_schematic}. 
The cathode is treated as the emitting boundary, where thermionic electrons are injected into the plasma and part of the emitted population may return to the cathode. 
The anode is treated as the downstream collection boundary, where transmitted emitted electrons are absorbed. 
These boundary fluxes define the imposed emitted-electron flux $\Gamma_{\mathrm{emit}}$, 
the cathode-directed backflow flux $\Gamma_{\mathrm{ref}}$, 
the net emitted-electron flux $\Gamma_{\mathrm{net}}=\Gamma_{\mathrm{emit}}-\Gamma_{\mathrm{ref}}$, 
and the anode-collected emitted-electron flux $\Gamma_{\mathrm{anode}}$.

The cathode-side energy balance is evaluated in two steps. 
First, the nominal energy carried away by thermionic emission is estimated. 
In ETC models, the energy removed by emitted electrons is commonly written as the sum of the work-function contribution and the average kinetic energy of emitted electrons 
\cite{hanquist_detailed_2017,boyer_mechanisms_2025}. 
Accordingly, the cathode-side emitted-electron cooling-power diagnostic is defined as
\begin{equation}
q_{\mathrm{emit,theory}}
=
\Gamma_{\mathrm{emit}} e
\left(W_{F,\mathrm{eff}}+2T_c\right),
\end{equation}
where $e$ is the elementary charge and $T_c$ is the imposed emission-electron temperature in eV. 
The term $2T_c$ represents the mean kinetic energy, in eV, carried by electrons emitted from a half-Maxwellian flux distribution.

Because the emitted-electron flux is imposed directly in the present simulations, 
an effective work function is inferred from a Richardson--Dushman-type thermionic-emission relation 
\cite{richardson_1903,dushman_electron_1923}:
\begin{equation}
W_{F,\mathrm{eff}}
=
-T_c \ln \left[
\frac{e\Gamma_{\mathrm{emit}}}{A_R T_c^2}
\right],
\end{equation}
where $A_R=1.20\times10^6/(k_B/e)^2$ is written in the eV-based temperature form. 
Therefore, $q_{\mathrm{emit,theory}}$ should be interpreted as an effective emission-energy diagnostic associated with the imposed flux, 
rather than as a material-specific prediction based on a fixed work function.

Second, the energy deposited back at the cathode is subtracted. 
The cathode-side heating terms include the kinetic-energy fluxes deposited by reflected emitted electrons, incident plasma electrons, and incident ions. 
Ion recombination heating is reported separately because ion recombination at a material surface can contribute to the wall-side energy balance 
\cite{monroeElectronTranspirationCircuits2025b,boyer_mechanisms_2025}. 
For helium, this contribution is estimated using the first ionization energy, $E_{\mathrm{ion}}=24.59~\mathrm{eV}$:
\begin{equation}
q_{\mathrm{ion,cath,recomb}}
=
\Gamma_{\mathrm{ion,cath}} e E_{\mathrm{ion}} .
\end{equation}

Two cathode-side cooling metrics are then defined. 
The first one subtracts the kinetic-energy deposition by reflected emitted electrons, plasma electrons, and ions:
\begin{equation}
q_{\mathrm{cooling},1}
=
q_{\mathrm{emit,theory}}
-
q_{\mathrm{back,em,cath}}
-
q_{\mathrm{back,p,cath}}
-
q_{\mathrm{ion,cath}} .
\end{equation}
The second one further subtracts the estimated ion-recombination heating at the cathode:
\begin{equation}
q_{\mathrm{cooling},2}
=
q_{\mathrm{cooling},1}
-
q_{\mathrm{ion,cath,recomb}} .
\end{equation}
With this sign convention, a larger positive value of $q_{\mathrm{cooling},1}$ or $q_{\mathrm{cooling},2}$ corresponds to a larger net cathode-side cooling diagnostic within the present diode model.

For completeness, a downstream transmitted-energy diagnostic is also evaluated from the kinetic-energy fluxes of emitted electrons, plasma electrons, and ions reaching the anode boundary:
\begin{equation}
q_{\mathrm{anode,total},1}
=
q_{\mathrm{out,em,anode}}
+
q_{\mathrm{out,p,anode}}
+
q_{\mathrm{ion,anode}} .
\end{equation}
The ion-recombination contribution at the anode boundary is estimated in the same way:
\begin{equation}
q_{\mathrm{ion,anode,recomb}}
=
\Gamma_{\mathrm{ion,anode}} e E_{\mathrm{ion}} .
\end{equation}
The recombination-included transmitted-energy diagnostic is therefore
\begin{equation}
q_{\mathrm{anode,total},2}
=
q_{\mathrm{anode,total},1}
+
q_{\mathrm{ion,anode,recomb}} .
\end{equation}

The corresponding boundary energy statistics are reported and discussed in Table~\ref{tab:energy_scan}.
The anode-side quantities are retained only to diagnose how much charged-particle energy reaches the downstream boundary in the benchmark diode. 
Because the present model is one-dimensional, these terms are not interpreted as the heat load on an actual finite-area collector with nonuniform electron landing positions. 
Accordingly, $q_{\mathrm{cooling},1}$ and $q_{\mathrm{cooling},2}$ should be interpreted as cathode-side boundary cooling metrics within the present kinetic diode model, 
not as complete material-level thermal balances for a hypersonic leading edge.

\FloatBarrier
\section{Results and discussion}

\subsection{Flux-scan results and transport-regime transition}

\begin{figure}[htbp]
\centering
(a)
\includegraphics[width=0.46\textwidth]{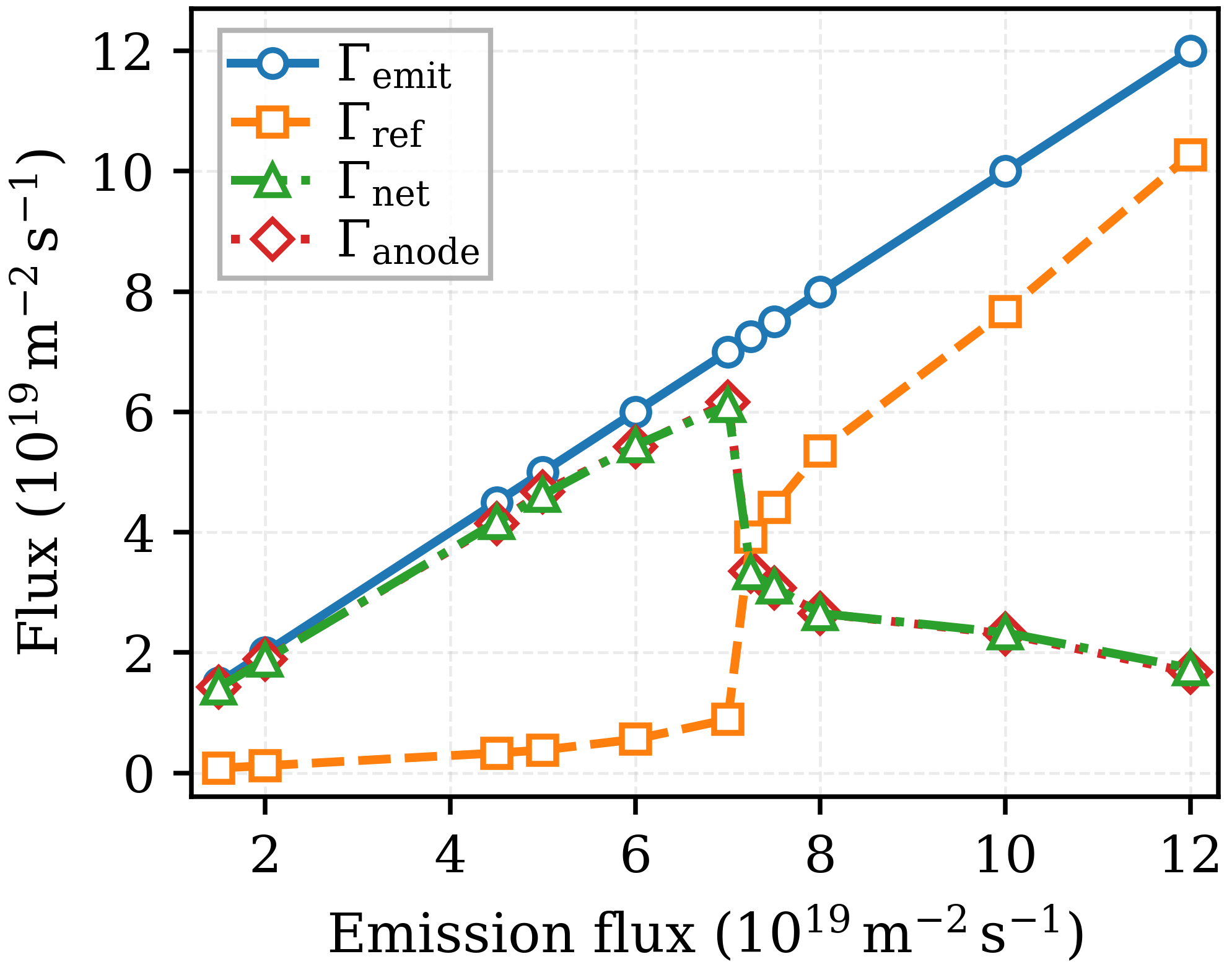}
(b)
\includegraphics[width=0.46\textwidth]{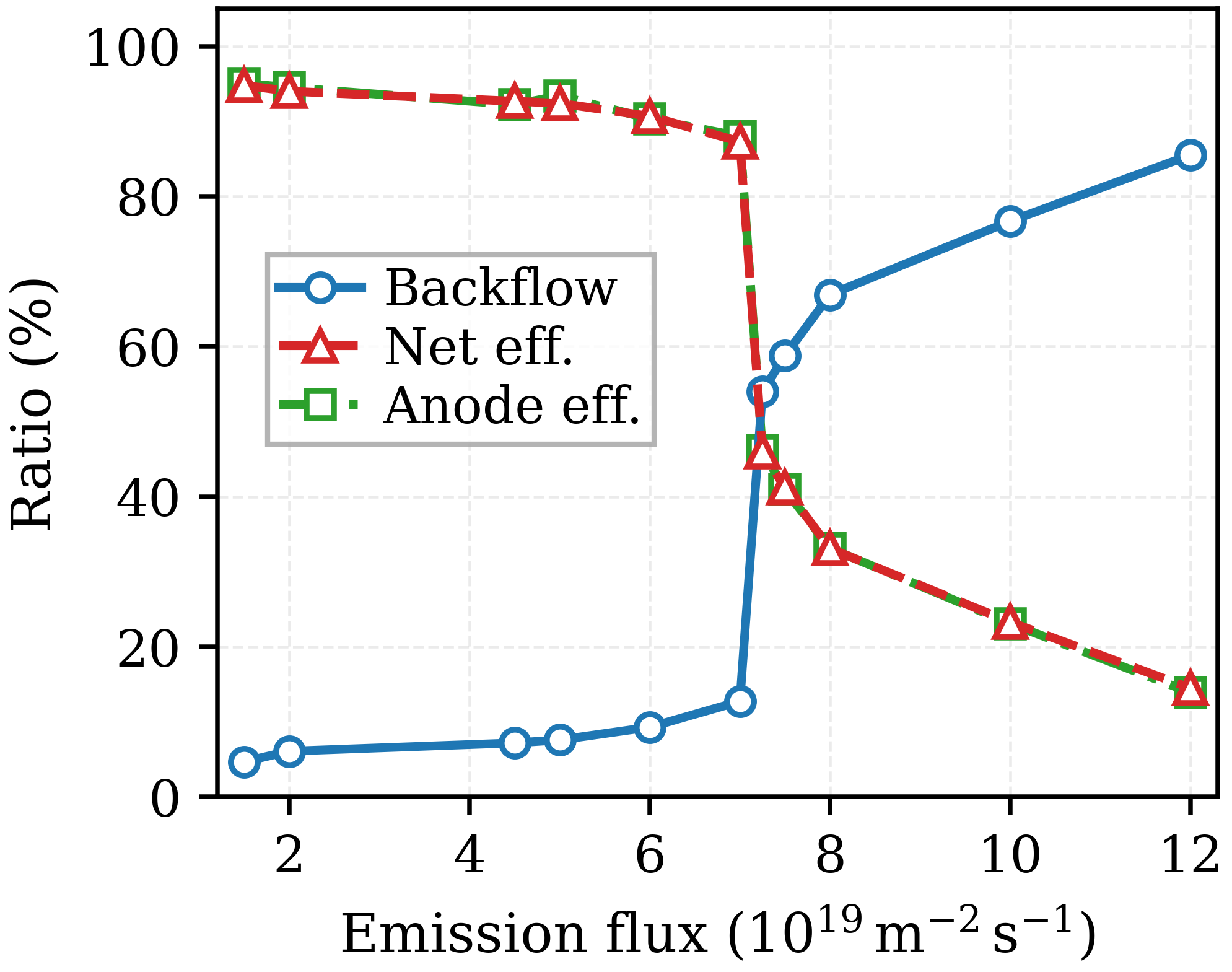}
\caption{
Emission-flux dependence of emitted-electron transport in the full-gap plasma diode.
(a) Imposed emission flux, cathode-directed backflow flux, net emitted-electron flux, and anode-collected emitted-electron flux.
(b) Backflow ratio, net transport efficiency, and anode collection efficiency.
The main transition from weak-backflow efficient-transmission transport to strongly backflow-limited transport occurs within
$\Gamma_{\mathrm{emit}}\approx(7.0$--$7.5)\times10^{19}\,\mathrm{m^{-2}\,s^{-1}}$.
The high-emission branch does not form an ideal plateau-like saturated current; instead, both $\Gamma_{\mathrm{net}}$ and $\Gamma_{\mathrm{anode}}$ decrease.
}
\label{fig:flux_transport_summary}
\end{figure}

\begin{table*}[htbp]
\centering
\caption{
Baseline flux data for the emission-flux scan under fixed collisional conditions.
All baseline cases use electron--neutral elastic scattering and ion--neutral charge exchange, without electron-impact excitation.
Fluxes are given in units of $10^{19}\,\mathrm{m^{-2}\,s^{-1}}$.
Here, $\Gamma_{\mathrm{emit}}$ is the imposed emitted-electron flux, $\Gamma_{\mathrm{ref}}$ is the cathode-directed backflow flux of emitted electrons, $\Gamma_{\mathrm{net}}=\Gamma_{\mathrm{emit}}-\Gamma_{\mathrm{ref}}$, and $\Gamma_{\mathrm{anode}}$ is the anode-collected emitted-electron flux.
}
\label{tab:benchmark_cases_baseline}
\footnotesize
\setlength{\tabcolsep}{4pt}
\renewcommand{\arraystretch}{1.12}
\begin{tabular}{ccccc}
\toprule
$\Gamma_{\mathrm{emit}}$ & $\Gamma_{\mathrm{ref}}$ & $\Gamma_{\mathrm{net}}$ & $\Gamma_{\mathrm{anode}}$ & Backflow ratio  \\
\makecell{($10^{19}\,\mathrm{m^{-2}\,s^{-1}}$)} &
\makecell{($10^{19}\,\mathrm{m^{-2}\,s^{-1}}$)} &
\makecell{($10^{19}\,\mathrm{m^{-2}\,s^{-1}}$)} &
\makecell{($10^{19}\,\mathrm{m^{-2}\,s^{-1}}$)} &
(\%)  \\
\midrule
1.50 & 0.08 & 1.42 & 1.43 & 5.21   \\
2.00 & 0.12 & 1.88 & 1.89 & 6.10  \\
4.50 & 0.33 & 4.17 & 4.15 & 7.20  \\
5.00 & 0.38 & 4.62 & 4.67 & 7.60  \\
6.00 & 0.56 & 5.44 & 5.42 & 9.30  \\
7.00 & 0.89 & 6.11 & 6.16 & 12.74  \\
7.25 & 3.92 & 3.33 & 3.35 & 54.03  \\
7.50 & 4.41 & 3.09 & 3.07 & 58.76  \\
8.00 & 5.35 & 2.65 & 2.65 & 66.91  \\
10.00 & 7.67 & 2.33 & 2.31 & 76.70  \\
12.00 & 10.27 & 1.74 & 1.67 & 85.50  \\
\bottomrule
\end{tabular}
\end{table*}

Fig.~\ref{fig:flux_transport_summary} summarizes the emission-flux scan, and Table~\ref{tab:benchmark_cases_baseline} lists the corresponding emitted-electron transport data averaged over selected late-time intervals. 
These intervals are chosen after the boundary fluxes fluctuate around statistically stable mean values.
As the imposed emission increases, the diode transitions from weak-backflow, efficient-transmission transport to strongly backflow-limited transport, with the main change occurring between $\Gamma_{\mathrm{emit}}=7.0\times10^{19}$ and $7.5\times10^{19}~\mathrm{m^{-2}\,s^{-1}}$.

At low emission, from $\Gamma_{\mathrm{emit}}=1.5\times10^{19}$ to $6.0\times10^{19}\,\mathrm{m^{-2}\,s^{-1}}$, most emitted electrons are transmitted across the gap. 
The reflected flux increases only from $0.08\times10^{19}$ to $0.56\times10^{19}\,\mathrm{m^{-2}\,s^{-1}}$, and the backflow ratio remains within $5.21\%$--$9.3\%$. 
Meanwhile, $\Gamma_{\mathrm{anode}}$ closely follows $\Gamma_{\mathrm{net}}$, indicating approximate emitted-electron transport balance in this regime.

The upper end of the efficient-transmission branch occurs near $\Gamma_{\mathrm{emit}}=7.0\times10^{19}\,\mathrm{m^{-2}\,s^{-1}}$. 
At this emission level, the reflected flux rises to $0.89\times10^{19}\,\mathrm{m^{-2}\,s^{-1}}$, the backflow ratio reaches $12.74\%$, and $\Gamma_{\mathrm{net}}$ and $\Gamma_{\mathrm{anode}}$ remain close to $6.1\times10^{19}\,\mathrm{m^{-2}\,s^{-1}}$. 
Thus, this case is best interpreted as an upper-end efficient-transmission or pre-transition reference state.

A pronounced transition occurs between $\Gamma_{\mathrm{emit}}=7.0\times10^{19}$ and $7.5\times10^{19}\,\mathrm{m^{-2}\,s^{-1}}$. 
At $\Gamma_{\mathrm{emit}}=7.25\times10^{19}\,\mathrm{m^{-2}\,s^{-1}}$, 
the late-time averaged reflected flux reaches 
$\Gamma_{\mathrm{ref}}=3.92\times10^{19}\,\mathrm{m^{-2}\,s^{-1}}$. 
Accordingly, $\Gamma_{\mathrm{net}}$ and $\Gamma_{\mathrm{anode}}$ are reduced to 
$3.33\times10^{19}$ and $3.35\times10^{19}\,\mathrm{m^{-2}\,s^{-1}}$, respectively.

The backflow ratio is therefore $54.03\%$, whereas the net transport and anode collection efficiencies are only about $46\%$. 
For $\Gamma_{\mathrm{emit}}\ge7.5\times10^{19}\,\mathrm{m^{-2}\,s^{-1}}$, the diode enters a strongly backflow-limited branch. 
The backflow ratio rises from $58.76\%$ at $7.5\times10^{19}\,\mathrm{m^{-2}\,s^{-1}}$ to $85.5\%$ at $1.2\times10^{20}\,\mathrm{m^{-2}\,s^{-1}}$, while $\Gamma_{\mathrm{net}}$ decreases from $3.09\times10^{19}$ to $1.74\times10^{19}\,\mathrm{m^{-2}\,s^{-1}}$. 
Thus, increasing the imposed emission no longer increases the useful transmitted flux; most of the additional emitted electrons are redirected back to the cathode.

The high-emission branch should not be interpreted as a failure to obtain current limitation. 
Rather, it represents a stronger form of full-gap transport limitation than an ideal plateau-like saturated branch. 
In a plateau-like backflow-saturated state, the increase in cathode-directed backflow approximately balances the increase in imposed emission, so that $\Gamma_{\mathrm{net}}$ remains nearly constant. 
Here, however, the backflow increases more rapidly than the imposed emission as $\Gamma_{\mathrm{emit}}$ is raised, causing both $\Gamma_{\mathrm{net}}$ and $\Gamma_{\mathrm{anode}}$ to decrease. 
This behavior is described as an overcompensated backflow-limited transport state.

This result remains consistent with the full-diode current-limitation framework of Campanell \emph{et al.}~\cite{PhysRevLett.134.145301}, in which emitted-current limitation can arise from backflow, space charge, or their cooperative interaction. 
The present results are best interpreted as a cooperative full-gap limitation in which cathode-side space-charge or virtual-cathode-like suppression and anode-side backflow limitation become coupled over a narrow emission interval.

As discussed in Sec.~2.2, matching the nominal mean-free-path ratio does not make the present 1D--3V MCC collision operator identical to the effective one-dimensional collisional relaxation used in the continuum benchmark. 
The resulting velocity-space redistribution can modify axial mobility and emitted-electron residence time, which helps explain why the high-emission branch decreases rather than forming an ideal plateau-like continuum branch.

\subsection{Full-gap potential restructuring under emission-flux scanning}

\begin{figure}[htbp]
\centering
(a)
\includegraphics[width=0.46\textwidth]{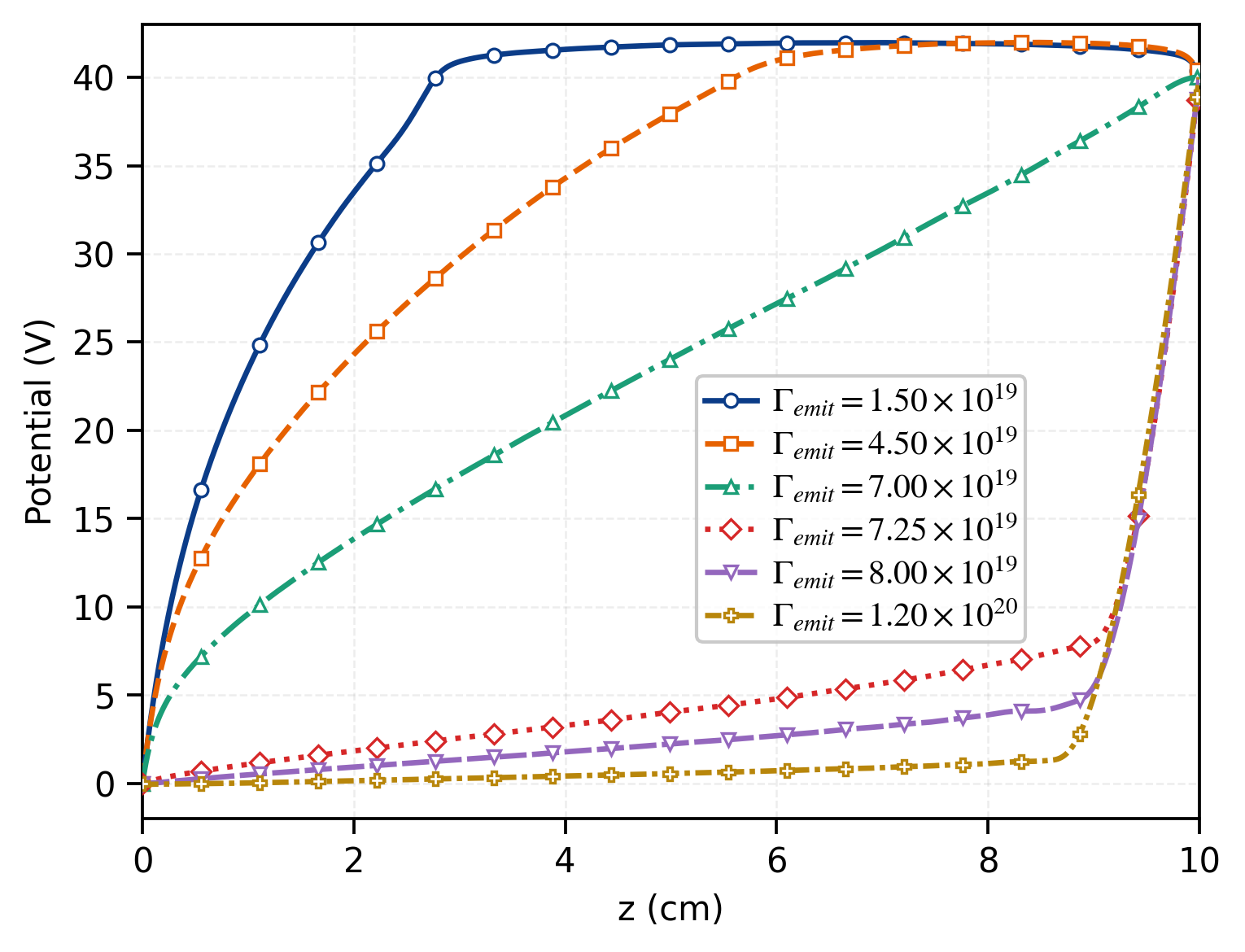}
(b)
\includegraphics[width=0.46\textwidth]{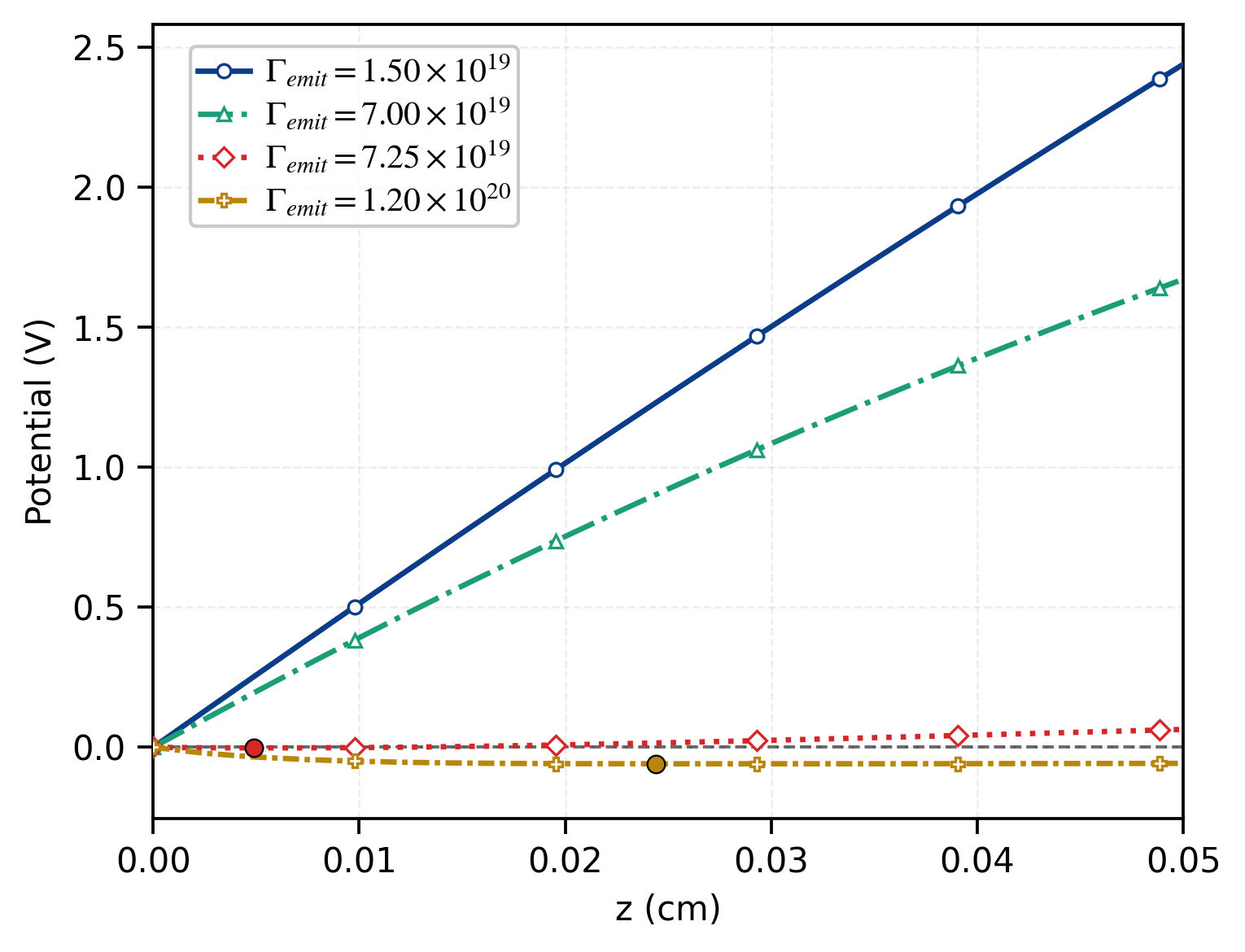}
\caption{
Late-time averaged potential profiles for selected representative baseline emission cases.
(a) Full-gap electrostatic potential profiles.
(b) Near-cathode zoom of the same profiles.
The filled circles in panel (b) mark local potential minima in the near-cathode region, which are used to assess the possible contribution of virtual-cathode-like reflection to emitted-electron return.
}
\label{fig:phi_scan}
\end{figure}

Fig.~\ref{fig:phi_scan} shows the late-time averaged potential restructuring for selected representative baseline emission cases. 
The full-gap profiles in Fig.~\ref{fig:phi_scan}(a) show that the transport transition identified in Fig.~\ref{fig:flux_transport_summary} is accompanied by a global reorganization of the electrostatic potential, while the near-cathode zoom in Fig.~\ref{fig:phi_scan}(b) resolves the local barrier weakening that promotes cathode-directed return motion. 
Thus, the degradation of emitted-electron transmission is not a purely local cathode-sheath effect, but a full-gap change of the diode state.

At low emission, represented by $\Gamma_{\mathrm{emit}}=1.5\times10^{19}\,\mathrm{m^{-2}\,s^{-1}}$, the potential rises rapidly near the cathode and remains close to a high-potential plateau over most of the gap, providing a favorable forward-transport environment. 
A similar structure remains at $4.5\times10^{19}\,\mathrm{m^{-2}\,s^{-1}}$, although the cathode-side rise becomes less abrupt. 
At $\Gamma_{\mathrm{emit}}=7.0\times10^{19}\,\mathrm{m^{-2}\,s^{-1}}$, the diode still transmits efficiently, but the full-gap potential is already lower and increases more gradually, consistent with its interpretation as a pre-transition state.

A much stronger restructuring occurs at $\Gamma_{\mathrm{emit}}=7.25\times10^{19}\,\mathrm{m^{-2}\,s^{-1}}$. 
The interior potential collapses substantially, remains low over most of the gap, and rises sharply only near the anode. 
This voltage redistribution explains why the high-emission branch does not remain at a constant transmitted-current plateau: once the plasma interior no longer provides an efficient forward-transport path, the enhanced backflow can suppress not only the additional emitted flux but also part of the previously transmitted emitted-electron population.

The near-cathode profiles further show that the local barrier against cathode-directed return motion is weakened in the transition-range and strongly backflow-limited cases. 
For the highest-emission case shown, a shallow negative dip appears near the cathode, indicating a virtual-cathode-like potential depression. 
However, this local feature is too weak to explain the sharp increase in reflected flux by itself.

Once the emission reaches $\Gamma_{\mathrm{emit}}=8.0\times10^{19}\,\mathrm{m^{-2}\,s^{-1}}$ and above, the interior potential remains much lower than in the low-emission regime, and the voltage rise is concentrated mainly near the anode. 
This confirms that emitted-electron current limitation in the present ETC-oriented diode is controlled by coupled cathode-sheath weakening, interior-potential collapse, and anode-side voltage redistribution.

To connect the present full-gap transport results with local emissive-sheath theory, the simulated net transport efficiency is compared with a virtual-cathode transmission estimate. 
In space-charge-limited or virtual-cathode sheath descriptions, the transmitted electron flux is exponentially sensitive to the cathode-side potential barrier
\cite{takamuraSpaceChargeLimitedCurrent2004a,jinInvestigationInfluenceMechanism2024}:
\begin{equation}
\eta_{\mathrm{VC}}
=
\frac{\Gamma_{\mathrm{VC}}}{\Gamma_{\mathrm{emit}}}
=
\exp\left[
\frac{\phi_{\mathrm{vc}}-\phi_{\mathrm{w}}}{T_{\mathrm{emit}}}
\right],
\end{equation}
where \(T_{\mathrm{emit}}\) is expressed in eV. 
Since \(\phi_{\mathrm{w}}=0\) and \(T_{\mathrm{emit}}=0.2\,\mathrm{eV}\), the global net transport efficiency can be mapped to an equivalent barrier potential,
\begin{equation}
\phi_{\mathrm{vc,eff}}
=
T_{\mathrm{emit}}
\ln
\left(
\frac{\Gamma_{\mathrm{net}}}{\Gamma_{\mathrm{emit}}}
\right).
\end{equation}

For $\Gamma_{\mathrm{emit}}=7.25\times10^{19}\,\mathrm{m^{-2}\,s^{-1}}$, $\Gamma_{\mathrm{net}}/\Gamma_{\mathrm{emit}}\approx0.46$, giving $\phi_{\mathrm{vc,eff}}\approx-0.156\,\mathrm{V}$. 
By contrast, the directly extracted near-cathode minimum in the averaged potential profile is only about $-2.2\times10^{-3}\,\mathrm{V}$, corresponding to $\eta_{\mathrm{VC}}\approx0.99$. 
Therefore, the strong transport loss cannot be explained by a local near-cathode virtual-cathode barrier alone. 
The much deeper $\phi_{\mathrm{vc,eff}}$ should instead be interpreted as an effective full-gap transport-loss parameter that includes interior-potential collapse, collisional redistribution, emitted-electron backflow from the plasma interior, and anode-side transport limitation.

\subsection{Temporal evolution near the transition range}

The temporal evolution is examined using the $\Gamma_{\mathrm{emit}}=7.25\times10^{19}\,\mathrm{m^{-2}\,s^{-1}}$ case, because this case lies inside the narrow transition interval and exhibits the clearest late-time restructuring behavior. 
As discussed in Sec.~3.1, this case is not merely an upper-end efficient-transmission state; instead, it already belongs to a strongly backflow-affected transition-range state whose full-gap structure continues to evolve over the accessible simulation window.

\begin{figure}[htbp]
\centering
\includegraphics[width=0.46\textwidth]{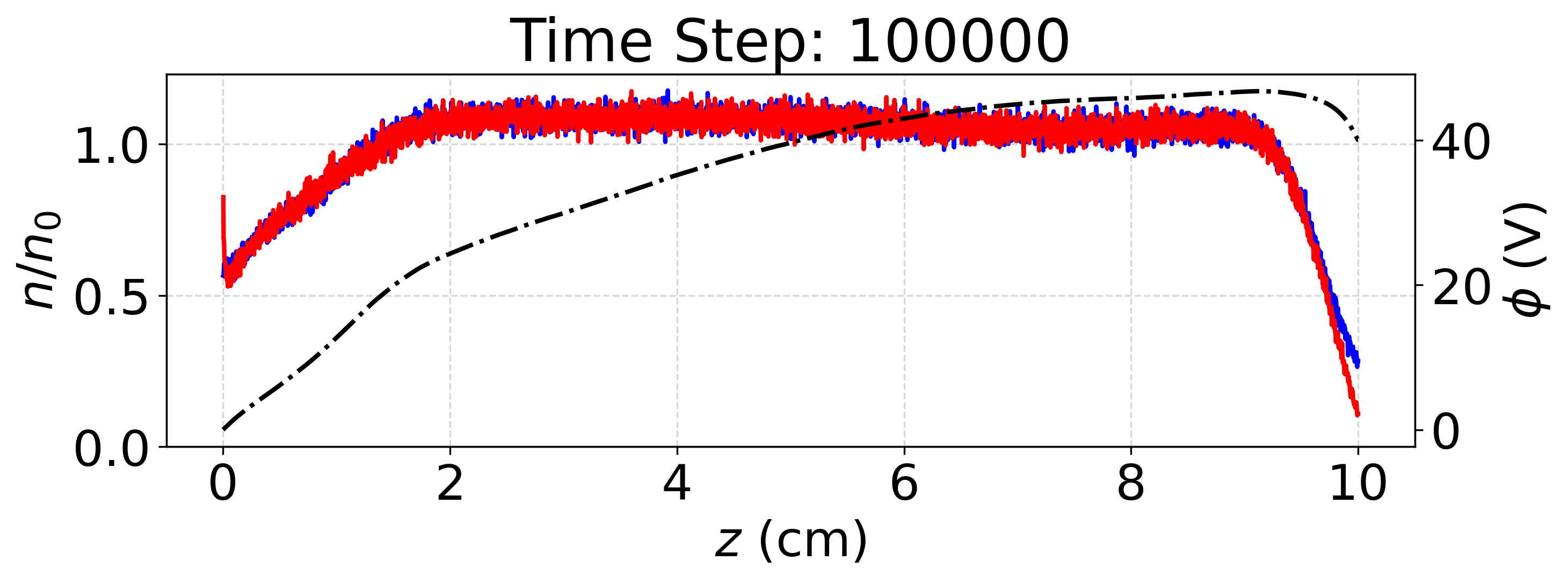}
\includegraphics[width=0.46\textwidth]{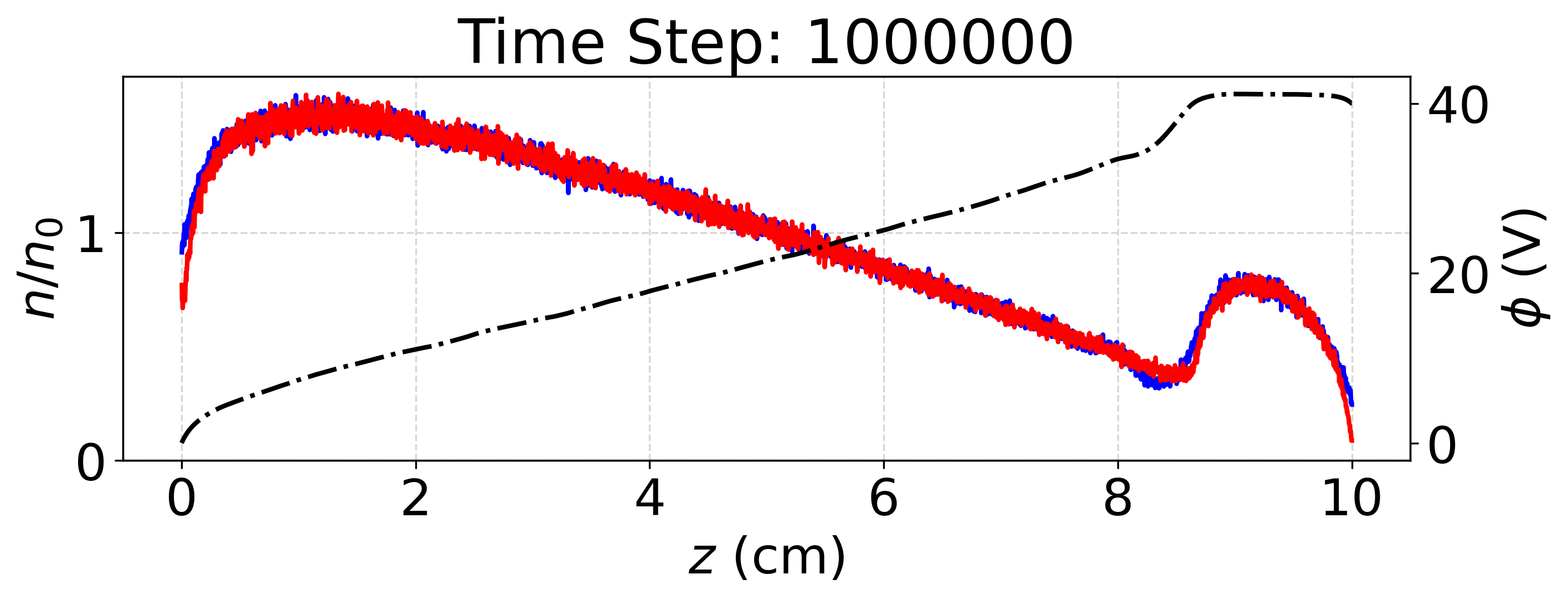}
\includegraphics[width=0.46\textwidth]{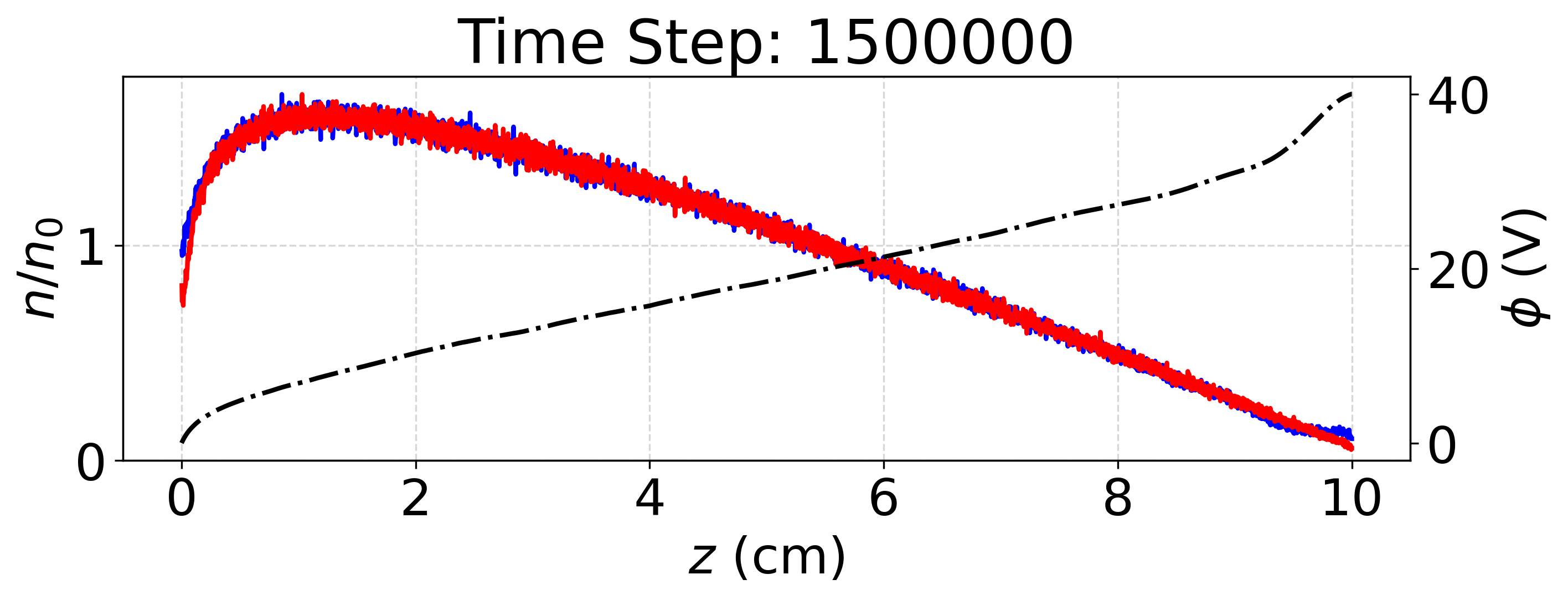}
\includegraphics[width=0.46\textwidth]{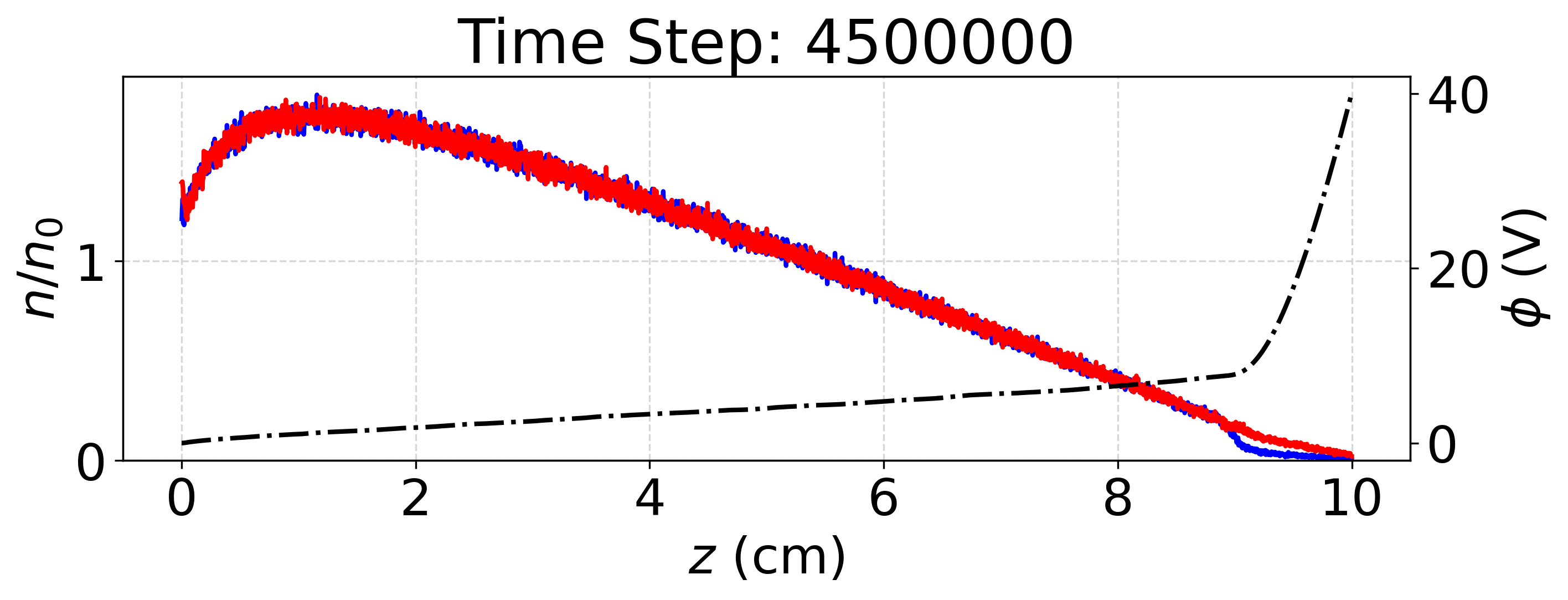}
\caption{
Particle-density and potential evolution for the transition-range case
$\Gamma_{\mathrm{emit}}=7.25\times10^{19}\,\mathrm{m^{-2}\,s^{-1}}$.
The panels correspond to $1.0\times10^{5}$, $1.0\times10^{6}$, $1.5\times10^{6}$, and $4.5\times10^{6}$ time steps.
In each panel, the blue solid curve denotes the normalized ion density $n_{\mathrm{i}}/n_0$, the red solid curve denotes the normalized total-electron density $n_{\mathrm{e}}/n_0$, and the black dashed curve denotes the electrostatic potential $\phi(z)$ referenced to the right vertical axis.
The sequence shows how the transition-range case evolves from an initially efficient-transmission configuration toward a strongly backflow-dominated transport state, together with progressive full-gap potential restructuring.
}
\label{Potential_vs_time_case2}
\end{figure}

Fig.~\ref{Potential_vs_time_case2} presents representative snapshots of the density--potential structure at four stages of the evolution. 
At early time, the diode still shows an efficient-transmission configuration, with the interior potential remaining comparatively favorable for forward emitted-electron transport.
As the run proceeds, the potential profile restructures globally, the interior potential decreases, and the cathode-side barrier weakens substantially. 
At the same time, the density distribution becomes increasingly asymmetric across the gap, with a stronger density gradient developing between the cathode side and the anode side. 
By the latest time shown, the diode has clearly departed from the initial efficient-transmission state and has evolved toward a strongly backflow-dominated transport configuration.

The late-time data for this case are therefore interpreted as quasi-steady transport averages. 
The averaging window is selected from the late-time interval in which the emitted, reflected, net, and collected emitted-electron fluxes fluctuate around statistically stable mean values. 
Thus, the reported fluxes characterize the late-time transport state rather than an instantaneous transient configuration.

\subsection{Selected sensitivity to collisional modeling details}

Table~\ref{tab:comparison_cases} summarizes selected sensitivity cases. 
Electron-impact excitation reduces backflow at low and intermediate emission: the backflow ratio decreases from $5.21\%$ to $3.30\%$ at $\Gamma_{\mathrm{emit}}=1.5\times10^{19}\,\mathrm{m^{-2}\,s^{-1}}$, from $7.20\%$ to $3.70\%$ at $4.5\times10^{19}\,\mathrm{m^{-2}\,s^{-1}}$, and from $7.60\%$ to $4.70\%$ at $5.0\times10^{19}\,\mathrm{m^{-2}\,s^{-1}}$. 
At high emission, however, excitation does not remove the backflow-limited state: at $8.0\times10^{19}\,\mathrm{m^{-2}\,s^{-1}}$, the backflow ratio is reduced from $66.91\%$ to $55.50\%$, while at $1.0\times10^{20}\,\mathrm{m^{-2}\,s^{-1}}$ the excitation-on case remains strongly backflow-limited with a backflow ratio of $80.80\%$. 
Increasing the electron-side neutral target temperature to $1740~\mathrm{K}$ at $\Gamma_{\mathrm{emit}}=7.0\times10^{19}\,\mathrm{m^{-2}\,s^{-1}}$ changes the backflow ratio only modestly, from $12.74\%$ to $11.60\%$. 
Disabling ion charge exchange at $\Gamma_{\mathrm{emit}}=1.0\times10^{20}\,\mathrm{m^{-2}\,s^{-1}}$ partially relieves the restriction, decreasing the backflow ratio from $76.70\%$ to $70.90\%$. 
Finally, increasing the particle number from $N_{\mathrm{ppc}}=1000$ to $2000$ changes the high-emission transport quantities only weakly. 
These comparisons show that the backflow-limited branch is quantitatively sensitive to collision details but is not caused by a single collisional or numerical-modeling choice.

\begin{table*}[!htbp]
\centering
\caption{
Selected comparison cases for electron-impact excitation, ion charge exchange, electron-side neutral target temperature, and particle-number sensitivity.
Fluxes are given in units of $10^{19}\,\mathrm{m^{-2}\,s^{-1}}$.
Here, Exc. denotes electron-impact excitation, and CX denotes ion--neutral charge exchange.
$\Gamma_{\mathrm{ref}}$ is the cathode-directed backflow flux of emitted electrons,
$\Gamma_{\mathrm{net}}=\Gamma_{\mathrm{emit}}-\Gamma_{\mathrm{ref}}$, and
$\Gamma_{\mathrm{anode}}$ is the anode-collected emitted-electron flux.
The backflow ratio is defined as
$\Gamma_{\mathrm{ref}}/\Gamma_{\mathrm{emit}}\times100\%$.
}
\label{tab:comparison_cases}
\footnotesize
\setlength{\tabcolsep}{3.5pt}
\renewcommand{\arraystretch}{1.12}
\begin{tabular}{cccccccc}
\toprule
Case & Exc. & CX & $\Gamma_{\mathrm{emit}}$ & $\Gamma_{\mathrm{ref}}$ & $\Gamma_{\mathrm{net}}$ & $\Gamma_{\mathrm{anode}}$ & Backflow  \\
 &  &  & \makecell{($10^{19}\,\mathrm{m^{-2}\,s^{-1}}$)} & \makecell{($10^{19}\,\mathrm{m^{-2}\,s^{-1}}$)} & \makecell{($10^{19}\,\mathrm{m^{-2}\,s^{-1}}$)} & \makecell{($10^{19}\,\mathrm{m^{-2}\,s^{-1}}$)} & (\%)  \\
\midrule
1  & Yes & On  & 1.50 & 0.05 & 1.46 & 1.42 & 3.30 \\
2  & Yes & On  & 4.50 & 0.17 & 4.33 & 4.34 & 3.70 \\
3  & Yes & On  & 5.00 & 0.24 & 4.77 & 4.75 & 4.70  \\
4  & Yes & On  & 8.00 & 4.44 & 3.56 & 3.57 & 55.50  \\
5  & Yes & On  & 10.00 & 8.09 & 1.92 & 1.92 & 80.80  \\
6 ($T_{\mathrm{n,e}}=1740\,\mathrm{K}$) & No & On & 7.00 & 0.81 & 6.18 & 6.26 & 11.60  \\
7  & No & Off & 10.00 & 7.09 & 2.91 & 2.83 & 70.90  \\
8 ($N_{\mathrm{ppc}}=2000$) & No & On & 10.00 & 7.66 & 2.34 & 2.31 & 76.60  \\
\bottomrule
\end{tabular}
\end{table*}

\FloatBarrier

\subsection{Cathode-side energy balance and ETC-relevant cooling}

Table~\ref{tab:energy_scan} lists late-time boundary energy-exchange diagnostics used to evaluate cathode-side energy balance and one-dimensional transmitted energy.

\begin{table*}[htbp]
\centering
\caption{Boundary energy-budget diagnostics for the available baseline emission cases. 
All quantities are in $\mathrm{W\,m^{-2}}$. 
The cathode-side terms quantify the effective emitted-electron cooling and the reheating contributions from reflected emitted electrons, plasma electrons, ions, and ion recombination. 
The anode-side terms are retained only as one-dimensional transmitted-energy diagnostics.}
\label{tab:energy_scan}
\footnotesize
\setlength{\tabcolsep}{3.2pt}
\renewcommand{\arraystretch}{1.12}
\begin{tabular}{cccccccccc}
\toprule
$\Gamma_{\mathrm{emit}}$ &
$q_{\mathrm{emit,theory}}$ &
$q_{\mathrm{back,em,cath}}$ &
$q_{\mathrm{back,p,cath}}$ &
$q_{\mathrm{ion,cath}}$ &
$q_{\mathrm{ion,cath,recomb}}$ &
$q_{\mathrm{cooling},1}$ &
$q_{\mathrm{cooling},2}$ &
$q_{\mathrm{anode,total},1}$ &
$q_{\mathrm{anode,total},2}$ \\
$(10^{19}\,\mathrm{m^{-2}\,s^{-1}})$ &
 & & & & & & & & \\
\midrule
1.50  & 14.72  & 0.07 & 0.03 & 12.06 & 12.02 & 2.56   & -9.47  & 86.83  & 91.43  \\
2.00  & 19.44  & 0.12 & 0.01 & 13.59 & 15.04 & 5.71   & -9.33  & 115.29 & 120.37 \\
4.50  & 42.57  & 0.33 & 0.06 & 20.71 & 27.11 & 21.46  & -5.65  & 258.75 & 264.28 \\
5.00  & 47.13  & 0.39 & 0.04 & 22.12 & 29.97 & 24.58  & -5.39  & 293.08 & 298.51 \\
7.00  & 65.23  & 0.87 & 0.10 & 15.65 & 33.94 & 48.61  & 14.67  & 406.57 & 406.76 \\
7.25  & 67.48  & 3.07 & 0.54 & 1.03  & 14.62 & 62.83  & 48.21  & 226.83 & 226.83 \\
7.50  & 69.72  & 3.29 & 0.61 & 0.74  & 12.83 & 65.08  & 52.25  & 207.67 & 207.67 \\
8.00  & 74.20  & 3.83 & 0.52 & 0.37  & 9.93  & 69.48  & 59.55  & 178.69 & 178.69 \\
10.00 & 92.04  & 5.30 & 0.50 & 0.23  & 7.78  & 86.01  & 78.23  & 147.75 & 147.75 \\
12.00 & 109.75 & 6.63 & 0.28 & 0.06  & 3.79  & 102.78 & 98.99  & 105.56 & 105.56 \\
\bottomrule
\end{tabular}
\end{table*}
Because the present model is one-dimensional, the anode-side energy terms are retained only as transmitted-energy diagnostics and are not interpreted as the heat load on an actual finite-area collector surface.
As the imposed emission increases, the cathode-side cooling estimate $q_{\mathrm{cooling},1}$ increases monotonically across the available baseline cases, from $2.56~\mathrm{W\,m^{-2}}$ at $\Gamma_{\mathrm{emit}}=1.5\times10^{19}~\mathrm{m^{-2}\,s^{-1}}$ to $48.61~\mathrm{W\,m^{-2}}$ at $7.0\times10^{19}~\mathrm{m^{-2}\,s^{-1}}$, $62.83~\mathrm{W\,m^{-2}}$ at $7.25\times10^{19}~\mathrm{m^{-2}\,s^{-1}}$, and more than $100~\mathrm{W\,m^{-2}}$ at $1.2\times10^{20}~\mathrm{m^{-2}\,s^{-1}}$. This trend means that, according to the present cathode-side boundary diagnostic, stronger emission removes more energy from the emitting surface.

The downstream transmitted-energy diagnostic behaves differently.
It first increases with emission and reaches its maximum near $\Gamma_{\mathrm{emit}}=7.0\times10^{19}~\mathrm{m^{-2}\,s^{-1}}$, where $q_{\mathrm{anode,total},2}$ is about $4.07\times10^{2}~\mathrm{W\,m^{-2}}$. 
After the diode enters the transition-range and strongly backflow-limited regimes, the downstream transmitted-energy diagnostic decreases
to about $2.27\times10^{2}~\mathrm{W\,m^{-2}}$ at $7.25\times10^{19}~\mathrm{m^{-2}\,s^{-1}}$, $1.79\times10^{2}~\mathrm{W\,m^{-2}}$ at $8.0\times10^{19}~\mathrm{m^{-2}\,s^{-1}}$, and $1.06\times10^{2}~\mathrm{W\,m^{-2}}$ at $1.2\times10^{20}~\mathrm{m^{-2}\,s^{-1}}$. 
Thus, increasing emission beyond the pre-transition state continues to enhance the nominal cathode-side cooling diagnostic, but it no longer improves useful full-gap emitted-electron transport. 
This behavior again shows that ETC-relevant performance is limited by emitted-electron escape and backflow, rather than by the imposed emission strength alone.

The stricter cooling metric $q_{\mathrm{cooling},2}$, which additionally subtracts cathode ion-recombination heating, gives a more conservative estimate of the cathode-side energy balance. Under this definition, the low- and intermediate-emission cases up to $\Gamma_{\mathrm{emit}}=5.0\times10^{19}~\mathrm{m^{-2}\,s^{-1}}$ still show net cathode-side heating, whereas the cooling balance becomes positive from the upper-end efficient-transmission regime onward. At $\Gamma_{\mathrm{emit}}=7.0\times10^{19}~\mathrm{m^{-2}\,s^{-1}}$, $q_{\mathrm{cooling},2}$ becomes positive and reaches about $14.67~\mathrm{W\,m^{-2}}$, and it continues to increase at higher emission.

These results indicate that ETC-relevant cathode cooling cannot be evaluated from the imposed emission strength alone. 
A larger emitted flux increases the nominal energy carried away by thermionic emission, but the effective cathode-side benefit also depends on the cathode-directed return of emitted electrons and on the plasma-electron, ion-impact, and recombination heating at the emitting boundary. 
Before the transport transition, increasing emission enhances both the cathode-side cooling diagnostic and the escape of emitted electrons from the cathode region. 
After the transition, however, stronger imposed emission continues to increase the nominal cathode-side cooling diagnostic but no longer improves useful full-gap emitted-electron transport, because an increasing fraction of the emitted population is redirected back to the cathode. 
This distinction is important for ETC because the cooling concept relies not only on electron emission from the hot surface, but also on the successful escape and transport of emitted electrons away from the cathode.

\subsection{Phase-space and velocity-distribution characteristics}

\begin{figure}[htbp]
\centering
\includegraphics[width=0.32\textwidth]{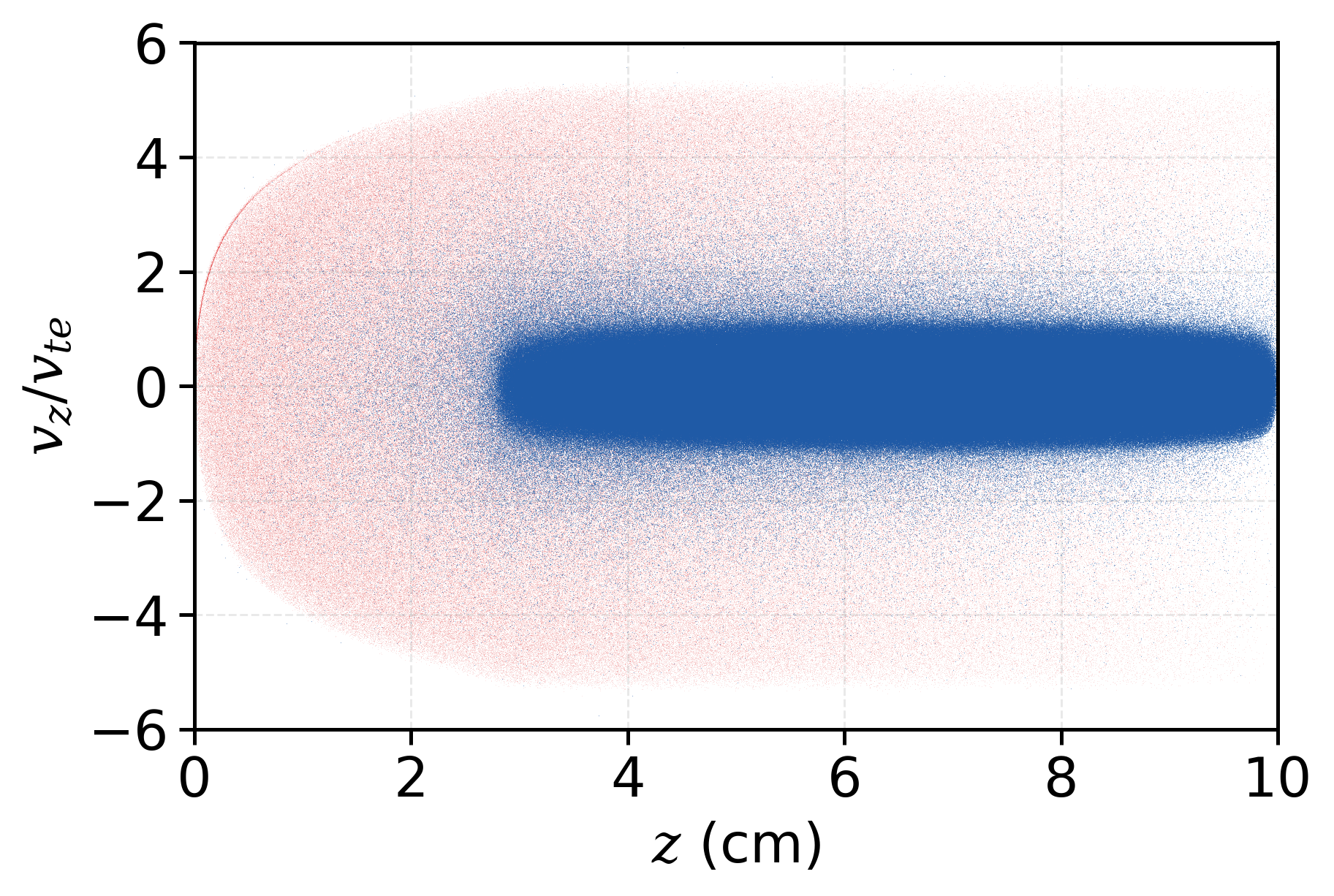}
\includegraphics[width=0.32\textwidth]{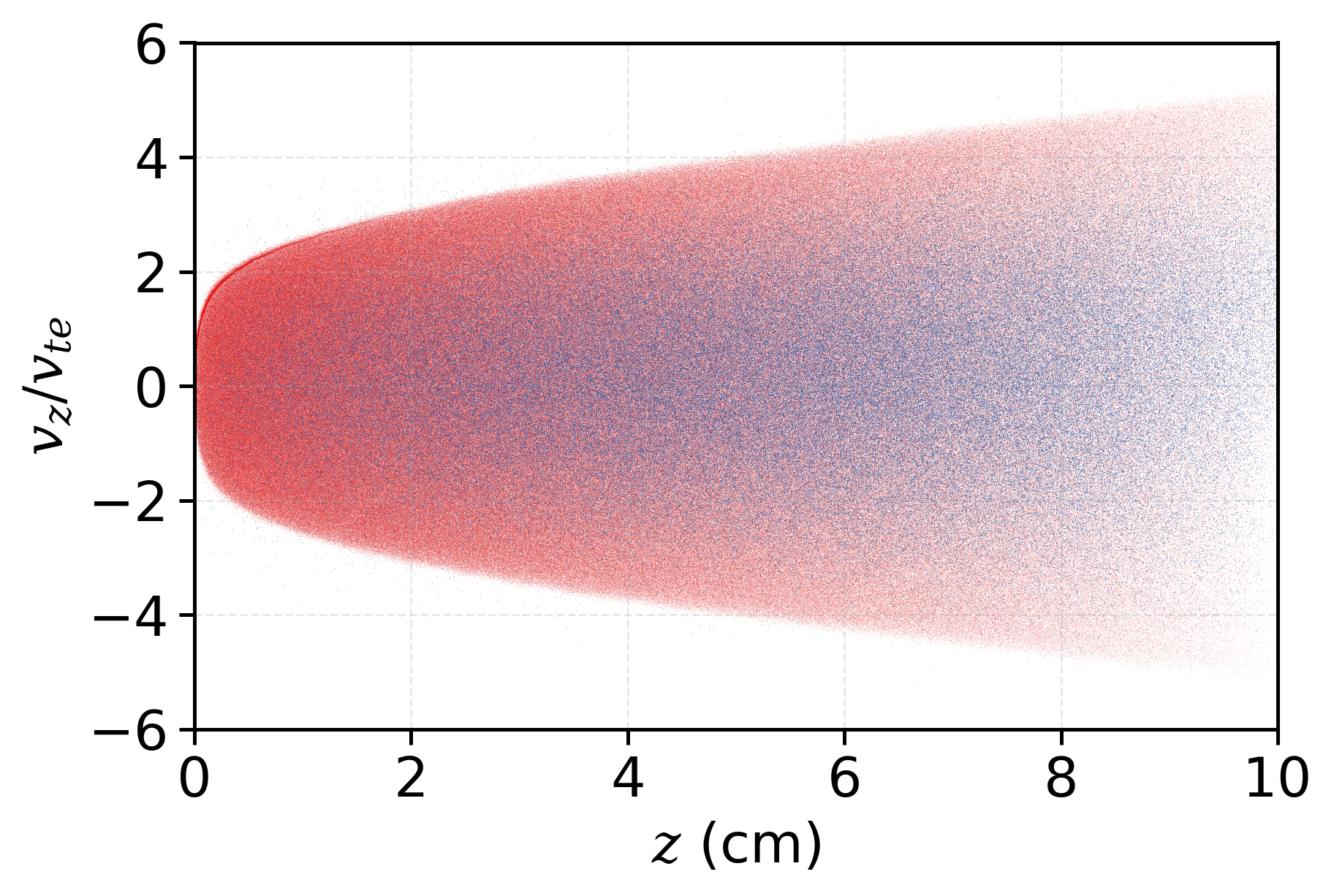}
\includegraphics[width=0.32\textwidth]{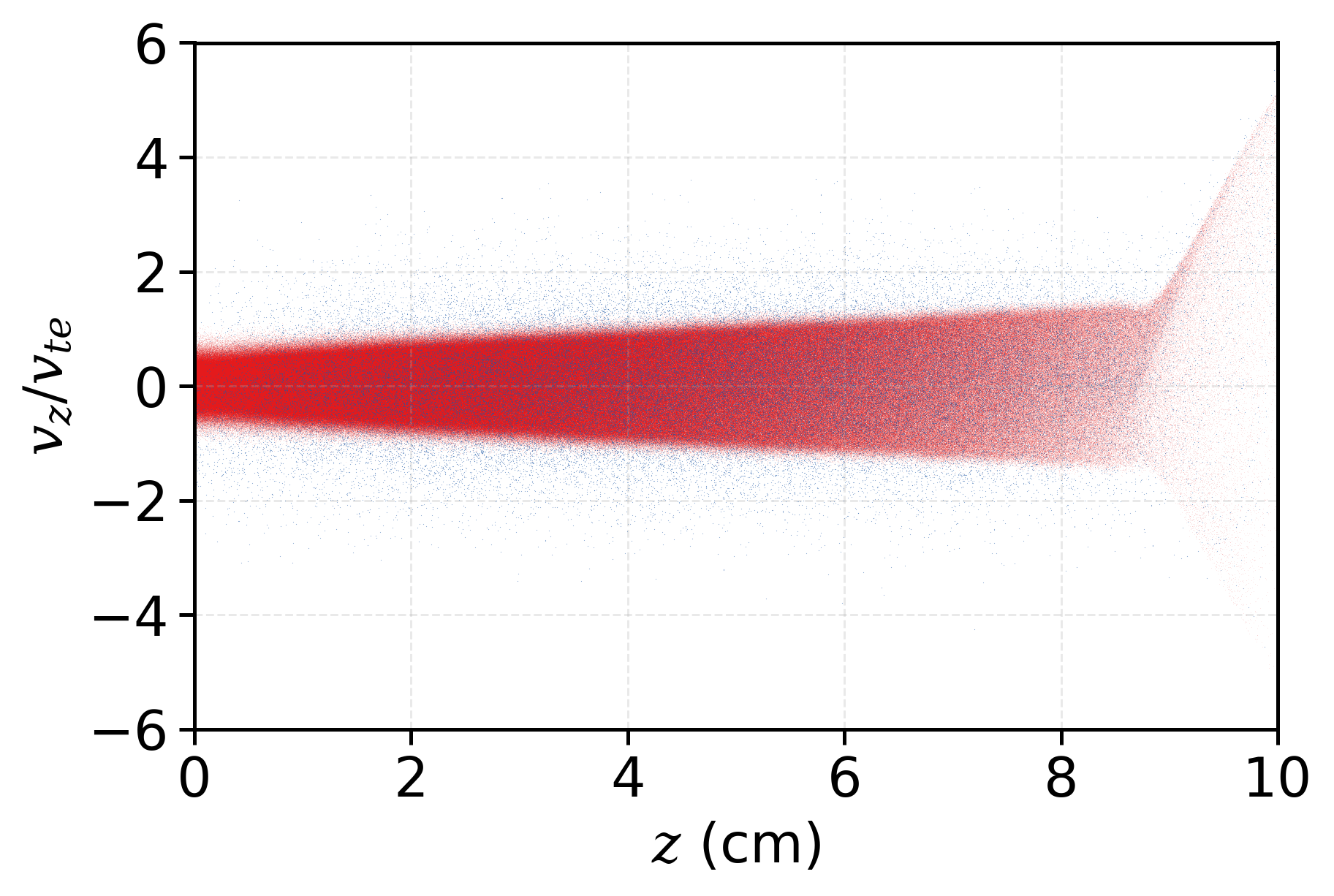}
\includegraphics[width=0.32\textwidth]{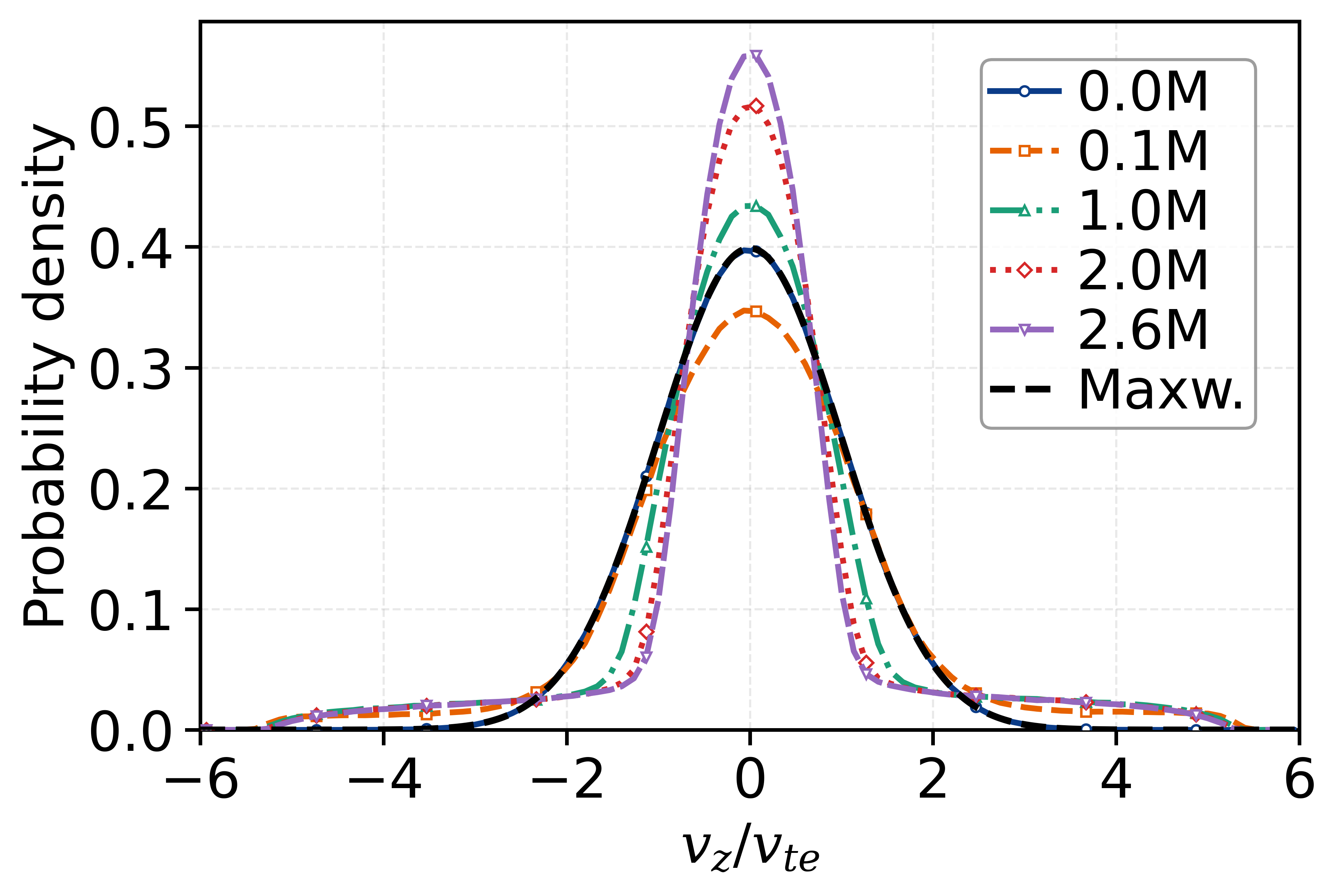}
\includegraphics[width=0.32\textwidth]{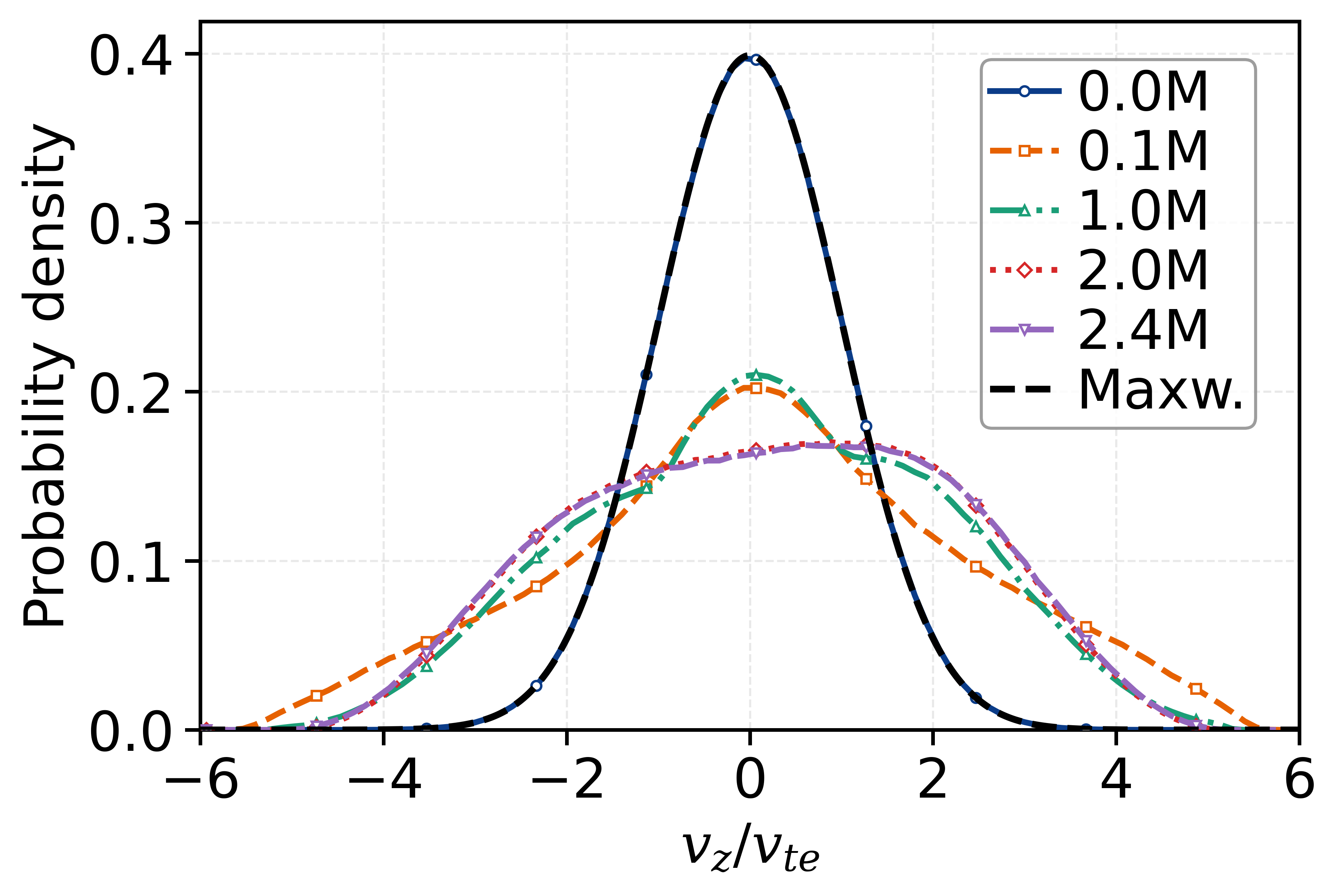}
\includegraphics[width=0.32\textwidth]{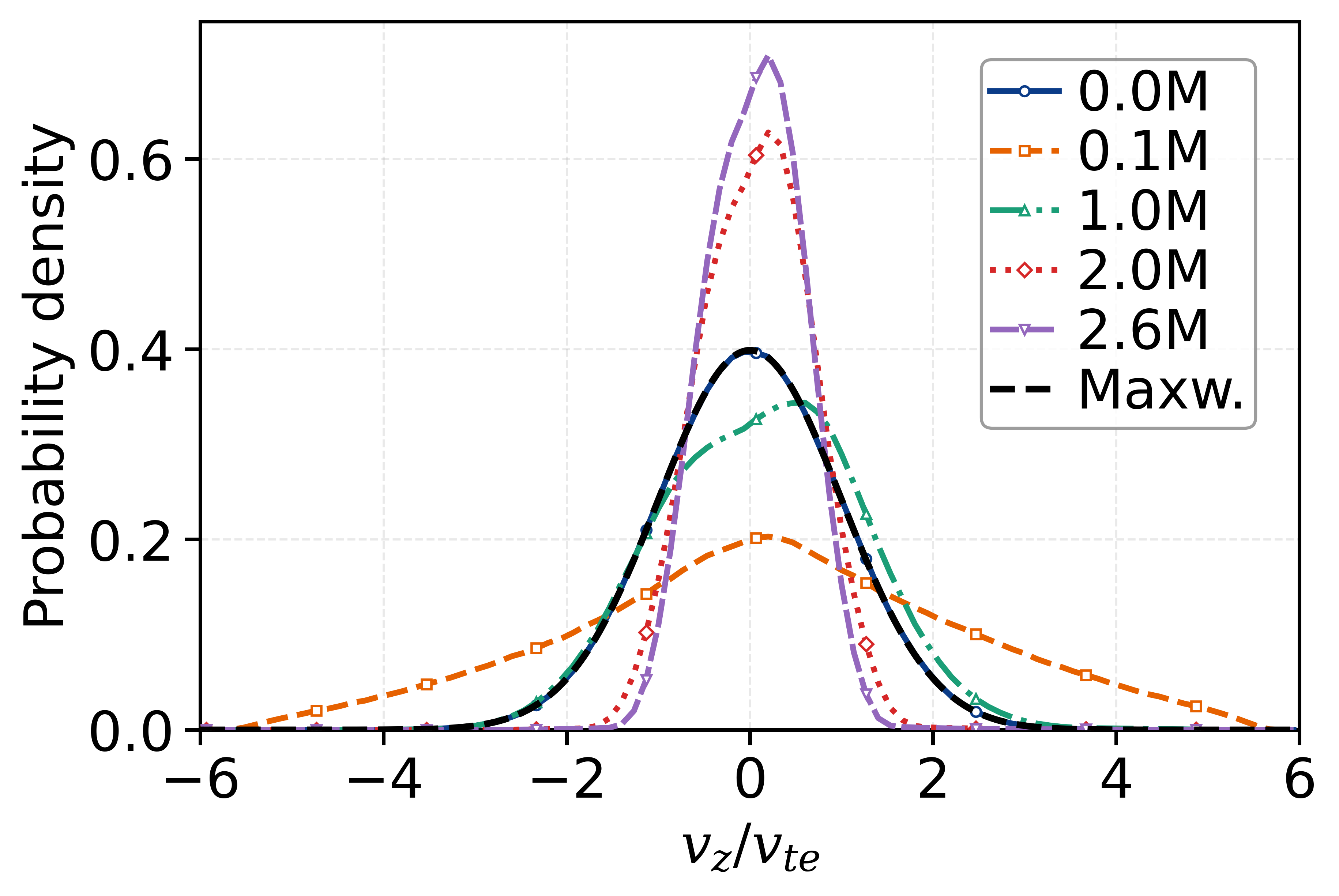}
\includegraphics[width=0.32\textwidth]{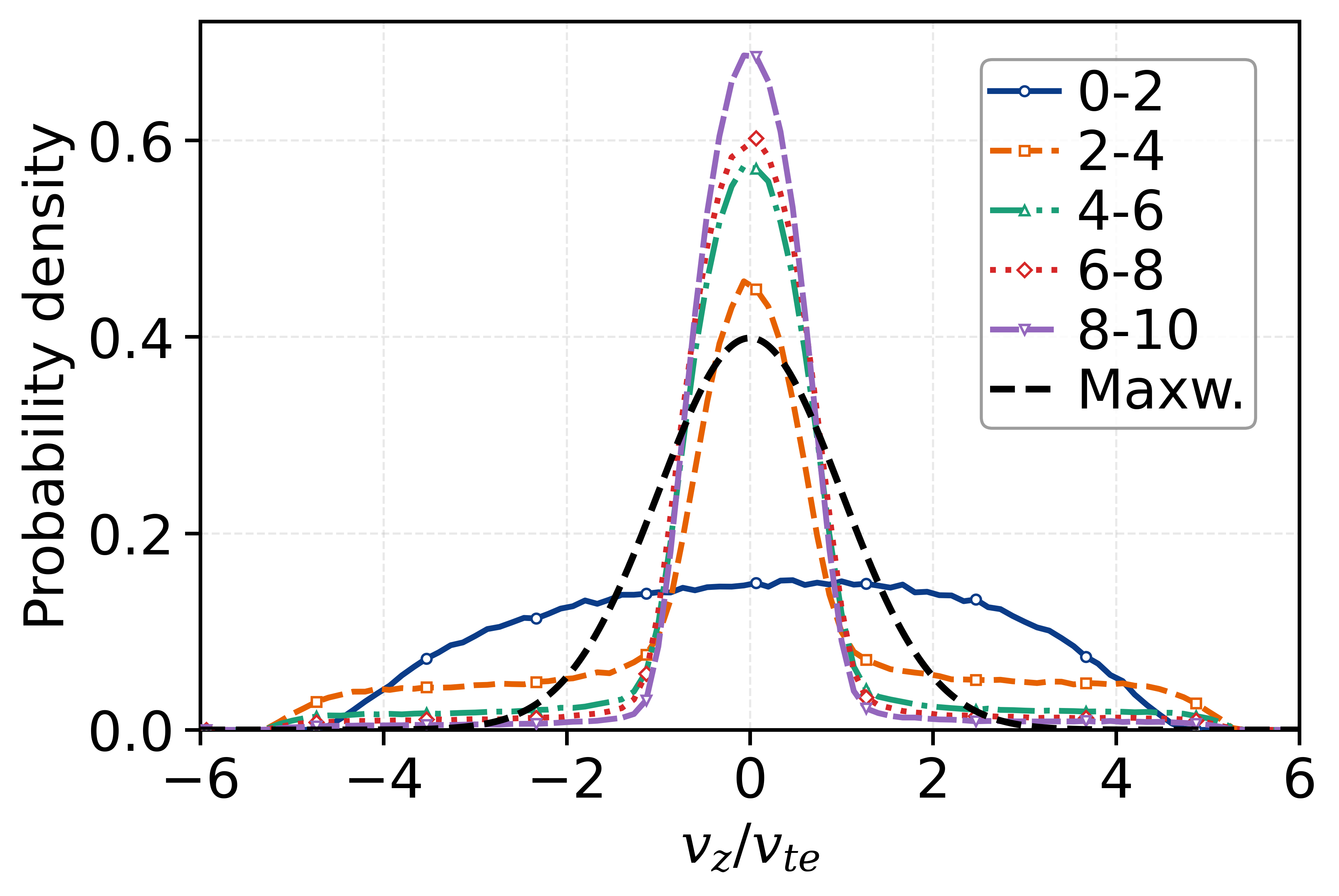}
\includegraphics[width=0.32\textwidth]{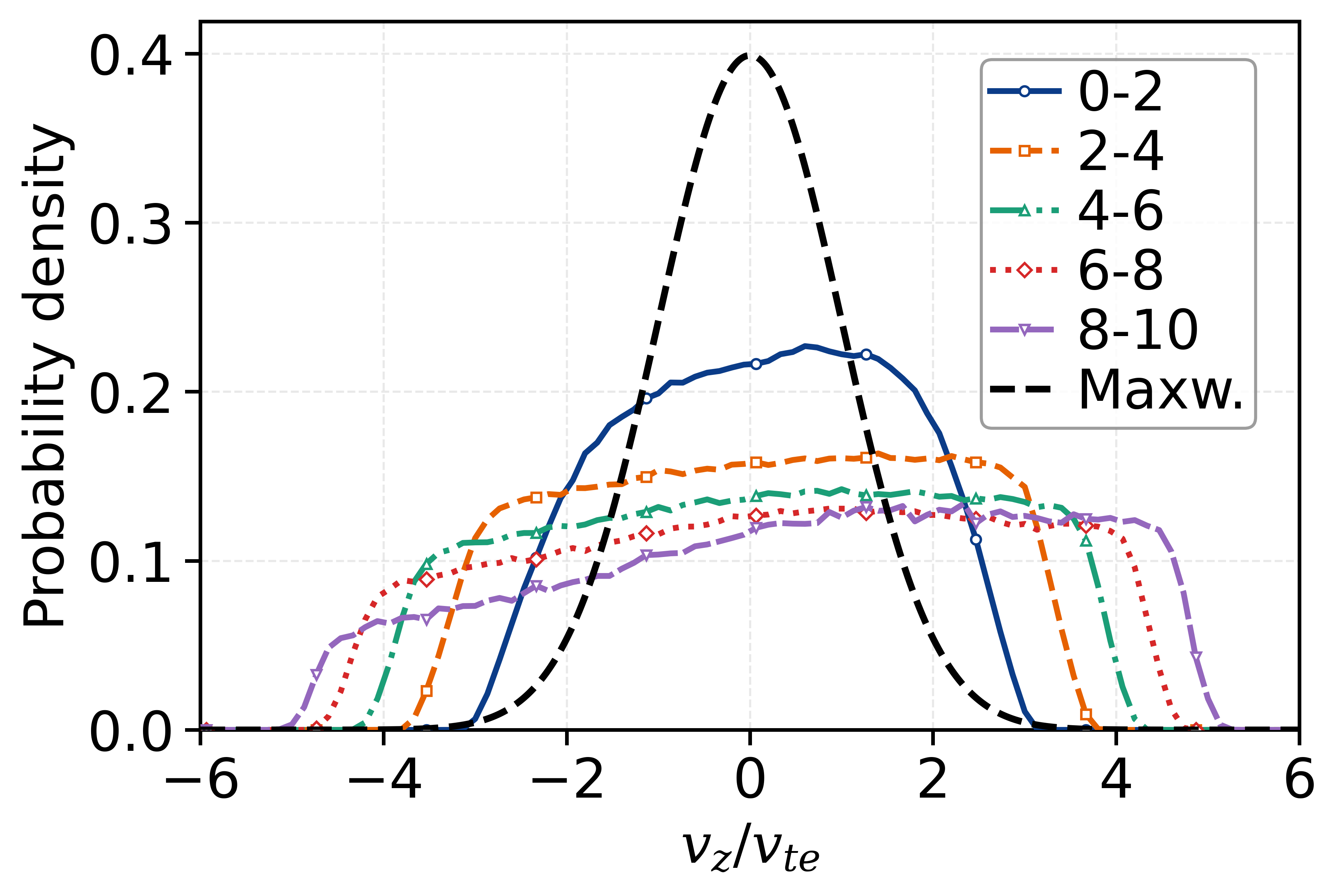}
\includegraphics[width=0.32\textwidth]{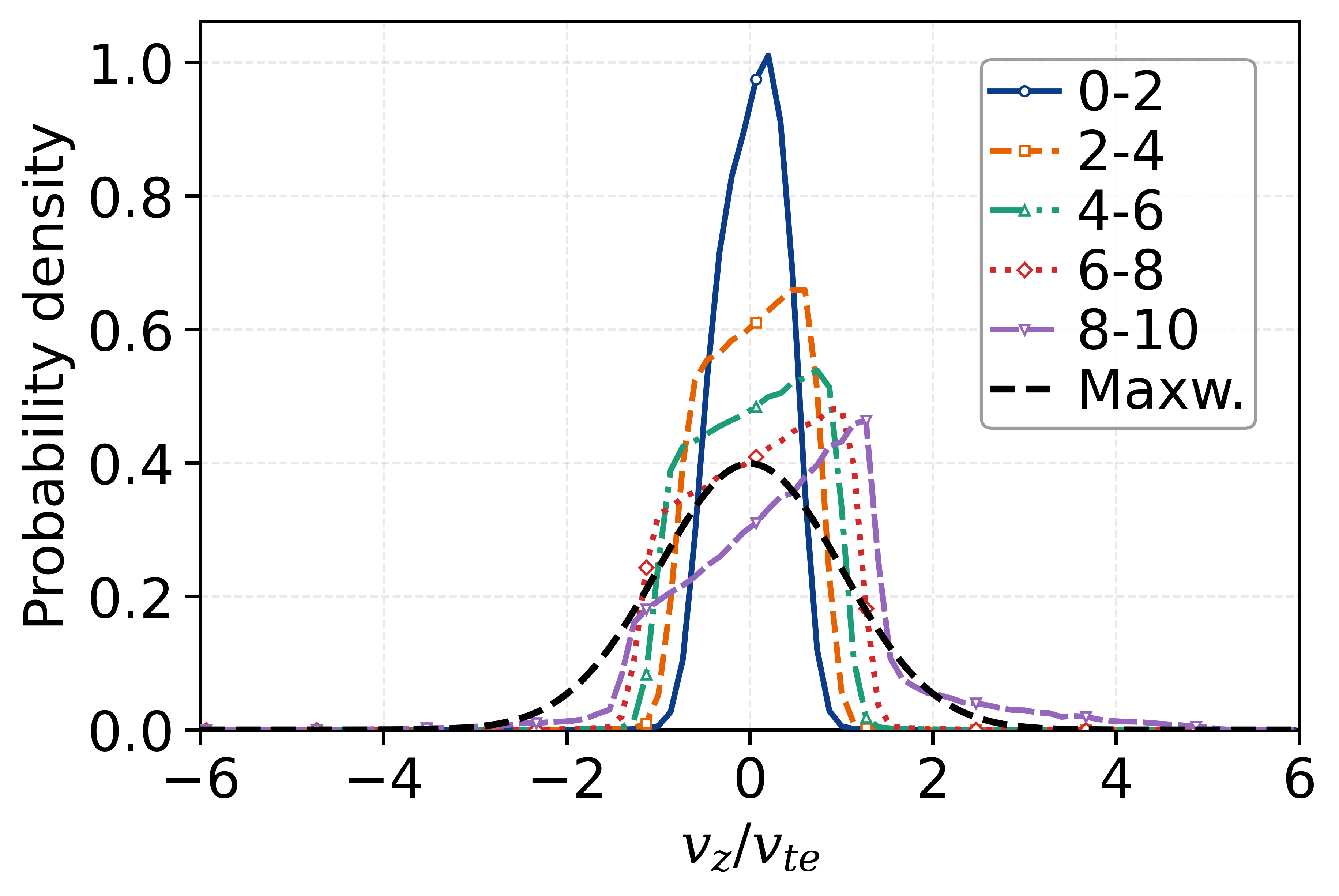}
\caption{
Electron phase-space and axial velocity-distribution diagnostics in three representative transport states.
The columns correspond to 
$\Gamma_{\mathrm{emit}}=1.5\times10^{19}$,
$7.0\times10^{19}$, and
$1.0\times10^{20}\,\mathrm{m^{-2}\,s^{-1}}$, respectively.
The upper row shows late-time phase-space distributions $(z,v_z)$, where emitted electrons and plasma electrons are plotted separately in red and blue.
The middle row shows the temporal evolution of the domain-integrated axial velocity distribution of all electrons; labels such as $0.1\mathrm{M}$ and $1.0\mathrm{M}$ denote simulation steps in millions.
The lower row shows late-time spatially resolved axial velocity distributions of all electrons; labels such as $0$--$2$ and $2$--$4$ denote sampling intervals in centimeters along the gap.
The dashed Maxwell curve denotes the reference Maxwellian distribution based on $T_{\mathrm{e},0}=3\,\mathrm{eV}$.
The velocity is normalized by $v_{\mathrm{te}}=(eT_{\mathrm{e},0}/m_{\mathrm e})^{1/2}$.
}
\label{fig:phase_space}
\end{figure}

Fig.~\ref{fig:phase_space} provides kinetic evidence for the transport-regime transition identified from the flux and potential diagnostics. 
The three columns represent a low-emission efficient-transmission case, an upper-end efficient-transmission/pre-transition case, and a strongly backflow-limited high-emission case. 
The upper panels distinguish emitted and plasma electrons in phase space, while the middle and lower panels diagnose the axial velocity distribution of the total electron population. 

In the low-emission case, $\Gamma_{\mathrm{emit}}=1.5\times10^{19}\,\mathrm{m^{-2}\,s^{-1}}$, the phase space is dominated by background plasma electrons, and the emitted-electron component occupies only a small fraction of the distribution. 
The domain-integrated axial velocity distribution remains close to the reference Maxwellian-like shape, and the spatially resolved distributions are similar across different gap regions. 
These features are consistent with weak backflow and efficient cross-gap transmission.

At $\Gamma_{\mathrm{emit}}=7.0\times10^{19}\,\mathrm{m^{-2}\,s^{-1}}$, the emitted-electron population becomes more extended in both space and velocity, and a cathode-directed component becomes more visible. 
The domain-integrated distribution broadens and flattens relative to the reference Maxwellian, indicating stronger kinetic mixing among emitted electrons, background plasma electrons, transmitted electrons, and reflected electrons. 
The spatially resolved distributions also become more region dependent, showing that the electron population has already been reorganized by full-gap transport, collisional scattering, and boundary losses even before the strongly backflow-limited branch is reached.

In the high-emission case, $\Gamma_{\mathrm{emit}}=1.0\times10^{20}\,\mathrm{m^{-2}\,s^{-1}}$, the emitted-electron component dominates over a large part of the gap, while cathode-directed motion is enhanced by the weakened near-cathode barrier and the backflow-limited transport state. 
The axial velocity distribution becomes strongly non-Maxwellian and reflects the coexistence of slow emitted electrons, transmitted electrons, reflected electrons, locally slowed electrons, and background plasma electrons. 
The strong regional variation of the late-time distributions confirms that the high-emission state is a kinetically mixed, backflow-limited state with low-velocity accumulation and nonlocal electron transport.

The velocity-distribution panels are therefore used as kinetic diagnostics, not as temperature-fitting plots. 
As emission increases, the total electron distribution evolves from a weakly perturbed Maxwellian-like shape to a broadened mixed distribution and finally to a strongly non-Maxwellian distribution with enhanced low-velocity accumulation. 
This behavior is expected in an open, boundary-driven diode with continuous cathode injection and absorbing electrodes. 
Electron--neutral elastic scattering redistributes velocity direction but does not force the total electron population to relax rapidly to a single Maxwellian equilibrium. 
This is also the kinetic origin of the difference between matching a nominal mean-free-path ratio and matching the full collision operator: the 1D--3V MCC model allows axial momentum to be redistributed into transverse motion, thereby modifying axial transport and supporting the interpretation of full-gap kinetic redistribution and backflow-limited transport.

The present benchmark cases do not aim to determine the onset condition of kinetic instabilities.
Nevertheless, the phase-space redistribution observed here suggests a possible route for future ETC studies.
When more realistic energy-dependent collision cross sections, lower neutral densities, or molecular gas mixtures are considered, the emitted-electron population may remain more beam-like over a longer distance before being collisionally isotropized.
In that case, relative drift between the emitted electrons and the background plasma electrons may excite beam--plasma or two-stream-like oscillations.
Such instabilities are expected to modify the full-gap potential fluctuation level, electron energy redistribution, residence time, and possibly the balance between transmitted and cathode-directed emitted electrons.
Therefore, a future extension of the present model should examine whether the ETC-relevant transport limitation changes from a mainly collisional backflow-limited state to an oscillatory instability-mediated transport state under lower-collisionality or air-like conditions.

\section{Conclusion}

A one-dimensional-in-space, three-dimensional-in-velocity electrostatic PIC-MCC model was developed to study the full emission--transport--backflow process of thermionically emitted electrons in a cathode--anode plasma diode relevant to electron transpiration cooling. 
The reported current data are interpreted as quasi-steady transport averages obtained from late-time intervals with statistically stable boundary fluxes. 
The main conclusions are summarized as follows.

\begin{enumerate}
\item 
Useful emitted-electron transport does not increase monotonically with imposed emission. 
A narrow transition occurs between 
$\Gamma_{\mathrm{emit}}=7.0\times10^{19}$ and 
$7.5\times10^{19}\,\mathrm{m^{-2}\,s^{-1}}$. 
At $\Gamma_{\mathrm{emit}}=7.25\times10^{19}\,\mathrm{m^{-2}\,s^{-1}}$, 
the backflow ratio reaches $54.03\%$, while the net transport and anode collection efficiencies decrease to about $46\%$. 
At higher emission, both $\Gamma_{\mathrm{net}}$ and $\Gamma_{\mathrm{anode}}$ decrease, indicating an overcompensated backflow-limited transport state.

\item 
The transition is caused by full-gap kinetic restructuring rather than by a purely local cathode-sheath effect. 
Potential profiles, density evolution, and phase-space diagnostics show interior-potential collapse, near-cathode barrier weakening, and strongly non-Maxwellian electron redistribution. 
Compared with the continuum benchmark of Campanell et al.~\cite{PhysRevLett.134.145301}, the present high-emission branch decreases instead of forming an ideal plateau-like branch 
because the 1D--3V MCC collision operator modifies axial momentum relaxation, emitted-electron residence time, and cathode-directed return probability.

\item 
Boundary energy statistics show that stronger imposed emission can continue to increase the nominal cathode-side boundary-cooling metric, even after useful full-gap emitted-electron transport no longer improves. 
Therefore, ETC performance cannot be evaluated from emission strength alone, but must account for emitted-electron escape, backflow, and full-gap kinetic transport.
\end{enumerate}

Future work should incorporate energy-dependent molecular cross sections, nitrogen/oxygen/air-like mixtures, multidimensional geometries, and finite-area collection effects. 
Another important extension is to determine whether lower-collisionality or air-like conditions can trigger beam--plasma or two-stream-like instabilities of the emitted-electron population and thereby modify potential fluctuations, energy redistribution, and useful emitted-electron transmission in ETC.

\section*{Acknowledgments}
The authors acknowledge the support from the National Natural Science Foundation of China (Grant No. 52472403 and U23B2033).

\section*{Data Availability}
The data that support the findings of this study are available from the corresponding author upon reasonable request.

\bibliographystyle{unsrt}
\bibliography{paper_refs}
\end{document}